\documentclass[floats,twocolumn,prb,aps]{revtex4}
\usepackage{graphicx}
\usepackage{dcolumn} 
\usepackage{amsmath}
\usepackage{exscale}
\usepackage{bm}      
\usepackage{color}
\usepackage{epsfig,psfrag,subfigure,subeqnarray}
\usepackage{amssymb}

\def\lan{\langle}
\def\ran{\rangle}
\def\va{\varepsilon}

\def\vk{{\bf k}}
\def\vK{{\bf K}}

\def\vr{{\bf r}}

\def\vQ{{\bf Q}}
\def\vq{{\bf q}}

\newcommand{\bd}{\begin{equation}}
\newcommand{\ed}{\end{equation}}
\newcommand{\be}{\begin{equation}}
\newcommand{\ee}{\end{equation}}
\newcommand{\bt}{\begin{split}}
\newcommand{\et}{\end{split}}

\newcommand{\bn}{\begin{align}}
\newcommand{\en}{\end{align}}

\newcommand{\bea}{\begin{eqnarray}}
\newcommand{\eea}{\end{eqnarray}}
\newcommand{\ba}{\begin{array}}
\newcommand{\ea}{\end{array}}
\newcommand{\nn}{\nonumber}

\begin{document}

\title{Cross-over from trion-hole to exciton-polaron\\ in $n$-doped semiconductor quantum wells}
\author{Yia-Chung Chang$^{1,2*}$}
\author{Shiue-Yuan Shiau$^{1,3}$, Monique Combescot$^{4}$}

\address{$^{1}$ Research Center for Applied Sciences, Academic Sinica,
Taipei, 11529 Taiwan}
\address{$^{2}$ Department of Physics, National Cheng Kung University, Tainan, 701 Taiwan}
\address{$^{3}$ Physics Division, National Center for Theoretical Sciences, Hsinchu, 30013, Taiwan}
\address{$^{4}$ Sorbonne Universit\'e, CNRS, Institut des NanoSciences de Paris, 75005-Paris, France}

\date{\today }
\begin{abstract}

We present a theoretical study of photo-absorption in $n$-doped two-dimensional (2D) and quasi-2D  semiconductors that takes into account the interaction of the photocreated exciton with Fermi-sea (FS) electrons through (i) Pauli blocking, (ii) Coulomb screening, and (iii) excitation of FS electron-hole pairs---that we here restrict to one. The system we tackle is thus made of one exciton plus zero or one FS electron-hole pair.  At low doping, the system ground state is predominantly made of a ``trion-hole"---a trion (two opposite-spin electrons plus a valence hole) weakly bound to a FS hole---with a small exciton component. As the trion is poorly coupled to photon,  the intensity of the lowest absorption peak is weak; it increases with doping, thanks to the growing exciton component, due to a larger coupling between 2-particle and 4-particle states. Under a further doping increase, the trion-hole complex is less bound  because of Pauli blocking by FS electrons, and its energy increases. The lower peak then becomes predominantly due to an exciton dressed by FS electron-hole pairs, that is, an exciton-polaron. As a result,  the absorption spectra of $n$-doped semiconductor quantum wells show two prominent peaks, the nature of the lowest peak turning from trion-hole to exciton-polaron under a doping increase.   Our work also nails down the physical mechanism behind the increase with doping of the energy separation between the trion-hole peak and the exciton-polaron peak, even before the anti-crossing,  as experimentally  observed.

\end{abstract}
%

\maketitle

\section{Introduction}

Optical spectra associated with  excitons in the presence of a Fermi sea (FS) in bulk\cite{Mahan} or quantum well\cite{Schmit1,Sanders,Schmit2,Hawrylak,Bauer,ZnSe,Kloch,Suris,Suris2,Kloch2,Koud} semiconductors have been a subject of great interest for decades. Depending on the doping concentration, the Coulomb interaction between the photocreated exciton and the FS electrons  can lead to various exotic complexes  that come from the dressing of the exciton by Fermi-sea  excitations,  {\it i.e.}, FS electron-hole pairs---the ``FS hole"  which corresponds to a missing electron in the doped conduction band, being fundamentally different from the valence hole making the exciton. Interestingly, at low doping, a bound state can emerge from the interaction of a trion (two opposite-spin electrons bound to a valence hole) and a FS hole, known as Suris tetron\cite{Suris2,Kloch2,Koud}. When the FS contains just one electron, this 4-particle complex reduces to the conventional $X^-$ trion because there is no other hole state for the FS hole to scatter into to possibly form a bound state with the trion through repeated interactions.\

Experiments on optical excitation spectra in $n$-doped II-VI quantum wells (QWs) show two peaks that are commonly associated with the negatively charged trion (X$^-$) and the exciton (X), separated by an energy that increases steadily with doping concentration\cite{Kloch,Suris,Suris2}. This increase in energy difference contradicts common physical understanding: Indeed, the trion binding should decrease when the doping increases because of the increase of Pauli blocking  and  the reduction of  Coulomb interaction due to screening by the doping electrons. Theoretical studies using  many-body Green functions have confirmed that absorption spectra should show two prominent peaks, separated by an energy difference that increases with doping\cite{Suris,Suris2,Kloch2,Koud}. Although the precise nature of these two peaks has not been established, it is clear that they  must come from many-body interactions between an exciton and FS electron-hole pairs.

Today, a considerable interest is devoted to these bound complexes that also appear in the photo-absorption spectra of $n$-doped 2D transition-metal dichalcogenides (TMDs)\cite{MAK2013,Sidler2017,Lin2014,Ganchev}. In these new 2D materials, excitons and trions have unusually strong bindings due to a strong reduction of the dielectric constant\cite{Courtade,Szyn,Wang,Qiu2013} (the trion binding can be as large as 30meV\cite{MAK2013,RossNC2013,Jones2013,Singh2014,Liu2DM2015,Zhu2015SR,YangACS2015}). Strong bindings, along with  large binding energy differences, render these materials quite suitable for studying the interplay between excitons and trions under a doping increase, for temperature as high as room temperature.  \

In a previous work, we have demonstrated the possibility of observing the implausible ``trion-polariton" in doped 2D semiconductors when the Fermi sea is spin-polarized\cite{Courtade}. In this work, we study the evolution  of the photo-absorption spectra in 2D and quasi-2D QWs of  III-V and II-VI semiconductors  when the doping increases. We show that at low doping, the lowest peak corresponds to a bound ``trion-hole", that is, the ground state of a trion and a FS hole. The higher peak corresponds to the exciton-polaron\cite{singularity}, that is, an exciton dressed by FS electron-hole pairs, which tends to an exciton when the doping goes to zero. While the energies of the trion-hole and the exciton-polaron would eventually cross when  the doping increases,  Coulomb interactions between them  prevent this crossing from happening by producing an anti-crossing. For higher doping, the lower peak is due to   the exciton-polaron and the higher peak due to the trion-hole. \

In addition, the exciton  encounters a cascade of  anti-crossings, at smaller dopings, with the {\it excited} trion-hole states that are distributed within a swath of energy of the order of the Fermi-sea energy (see Fig.~\ref{fig:8}). This  broadens the higher  peak.
As a result, the energy separation between the low and high peaks  increases not only
after the anti-crossing, but, surprisingly, also before the anti-crossing due to the multiple anti-crossings with the excited trion-hole states. This understanding provides a much sought-after explanation for  the counterintuitive experimental findings that the separation between the two peaks {\it always} increases as the doping increases, despite the fact that the binding of the trion with respect to the exciton reduces when the doping increases. At even higher doping, contributions to the exciton-polaron from multiple FS electron-hole pairs become sizable and ultimately lead to Fermi edge singularities \cite{singularity}. \

\begin{figure}[t]
\centering
 \includegraphics[trim=0.5cm 0cm 0.5cm 0cm,clip,width=3.2in]  {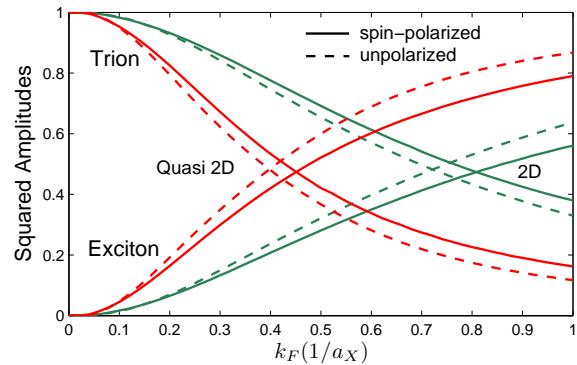}
\caption{\small  Squared amplitudes of the trion component and exciton component in the 4-particle ground state for spin-polarized (solid) and unpolarized (dashed) Fermi seas  as functions of the Fermi momentum $k_F$, for  2D  (green) and quasi-2D  (red) QWs.}
\label{fig:1}
\end{figure}

 To characterize the system ground state as a function of the Fermi momentum $k_F$, we show in Fig.~\ref{fig:1} the squared amplitudes of the trion-hole and exciton-polaron components of this ground state. For numerical purpose, the exciton-polaron considered here is dressed by zero or one FS electron-hole pair. Two cases are considered : (i) unpolarized FS, and (ii) polarized FS with electron spin  opposite to the spin of the photocreated electron.
  At the cross-over, the ground state has an equal amount of these two components. Before the cross-over, the ground state corresponds to a trion bound to a FS hole, whereas after the cross-over, the ground state leans predominantly toward exciton-polaron.
 With $a_X$ being the 3D exciton Bohr radius, the cross-over for spin-polarized (unpolarized) FS occurs at $k_Fa_X\simeq 0.8 ~(0.7)$ for 2D QW, but at $0.45 ~(0.4)$ for  quasi-2D GaAs/Al$_{0.25}$Ga$_{0.75}$ QW with 8nm thickness. This shows that Pauli blocking and Coulomb screening facilitate the transition to  exciton-polaron.\

To mathematically study the cross-over between trion-hole and exciton-polaron, we use two sets of basis functions: one set is made of one exciton (with its ground and excited levels) in the presence of a rigid FS; the other set is made of one exciton in the presence of one FS electron-hole pair; this basis  can also be taken as one trion (with its ground or excited level) with one FS hole. We perform a ``configuration-interaction" calculation to allow us to determine not only the ground state but also the excited states of the system. A particular focus is made on the Pauli exclusion principle induced by the  Fermi sea, either polarized or unpolarized, and its competition with the formation of the trion-hole complex  and exciton-polaron. What fundamentally distinguishes our method from the many-body Green's function method used in Ref.~\onlinecite{Suris2} is that our approach can provide a clear identification of the physical nature in each peak through the wave functions that are involved, an essential point to possibly study the cross-over.

To best understand the effect of Pauli blocking, we first study the binding energies of the exciton and trion in the presence of a ``frozen" Fermi sea, that is, a FS without any electron-hole excitation, when the exciton is photocreated in a semiconductor QW by the absorption of a circularly-polarized $(\sigma_+)$ photon; the resulting exciton is then made of a spin-$(-1/2)$ conduction electron and a $(+3/2)$ valence hole.
When the Fermi sea is unpolarized, both the exciton and the trion suffer Pauli blocking, which makes their binding energy decrease when the doping increases. When the Fermi sea is fully polarized with spin-$(1/2)$ electrons only, the photocreated exciton does not suffer Pauli blocking, whereas the spin-$(1/2)$ electron of the trion does. In both cases, we find that the lowest peak changes from trion-hole to exciton-polaron when the doping increases, but not at the same doping value.\

To tackle this complex system, we use the Rayleigh-Ritz variational method\cite{Ritz, MacDonald} that consists of expanding the bound states we want to determine on a finite set of localized basis functions. We solve the resulting Schr\"{o}dinger equation within this basis subspace, through a numerical diagonalization. The valence-hole effective mass is taken as infinite; this simplification, which does not affect the existence of bound states, allows reducing the basis functions to products of single-particle wave functions for electrons above the FS and hole in the FS. We have included the reduction of the Coulomb interaction induced by the doped electrons through the Thomas-Fermi screening.
To understand the effect of the QW thickness on this cross-over and catch the trend it induces, we have considered 2D and quasi-2D QWs by using an effective Coulomb scattering that varies from $1/q$ to $1/q^2$ when the well thickness increases and which have been shown to give excellent agreement with experiments\cite{Sanders}.

The paper is organized as follows. In Sec.~\ref{sec2}, we present the problem. In Sec.~\ref{sec3}, we introduce the basis functions that we use for the FS hole and the two electrons outside the FS. In Sec.~\ref{sec4} and \ref{sec5}, we calculate the low-lying exciton and trion states and their binding energies without and with a frozen FS, either fully polarized or unpolarized. The undoped case is used as a benchmark to show the accuracy of our variational procedure. In Sec.~\ref{sec6}, we turn to the trion-hole complex and the exciton-polaron, and study their cross-over. Next, we discuss the consequences on photo-absorption spectra, and we conclude.

\begin{figure}[t]
\centering
  \includegraphics[trim=2cm 5.5cm 1.5cm 6.9cm,clip,width=3.4in] {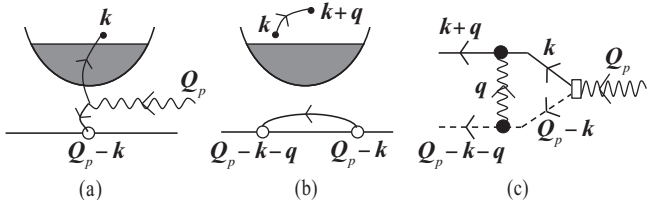}
\caption{\small (a) The absorption of a $\vQ_p$ photon creates a valence hole $(\vQ_p-\vk)$ and a conduction electron $\vk$ outside the FS. (b) Coulomb scattering of the photocreated pair for a frozen FS. Repeated interactions of the pair lead to the formation of an exciton, although weaker because some conduction states are Pauli blocked by the FS. (c) Feynman diagram for processes (a) and (b). Solid line represents conduction electron, dashed line represents valence hole. Squiggly line represents Coulomb interaction. }
\label{fig:2}
\end{figure}

\section{Physics of the problem\label{sec2}}

We  look for the absorption spectrum of a photon with momentum $\vQ_p$ in the presence of a Fermi sea that we take as $|F_N\ran=\prod_{i=1}^N a^\dag_{\vk_i}|v\ran$ with $|\vk_i|\leq k_F$, with $|v\ran$ denoting the vacuum state. To facilitate the argument, we  neglect the particle spin degrees of freedom; they will be introduced in due course.  The Fermi golden rule gives the photon absorption spectrum $\mathcal{A}(\vQ_p)$ through
\be
\mathcal{A}(\vQ_p)\propto \sum_f \big|\lan f|U^\dag_{ph-sc}|i\ran\big|^2\delta(E_f-E_i)\,,\label{AvQp}
\ee
where $|i\ran$ and $|f\ran$ are, respectively, the system eigenstates before and after photo-absorption, their energies being $E_i$ and $E_f$.

The initial state $|i\ran$ consists of the photon $\vQ_p$ with energy $\hbar\omega_{\vQ_p}$ and the Fermi sea $|F_N\ran$ with energy  $\mathcal{E}_N$, that is, $|i\ran=\alpha^\dag_{\vQ_p}|F_N\ran$ with energy  $E_i=\hbar\omega_{\vQ_p}+\mathcal{E}_N$. When  acting on $|i\ran$, the photon-semiconductor coupling $U^\dag_{ph-sc}$ destroys the photon $\vQ_p$ and creates a pair of conduction electron and valence hole with same total momentum (see Fig.~\ref{fig:2}(a)); so,
\be
U^\dag_{ph-sc}|i\ran=\Omega \sum_{\vk}a^\dag_\vk b^\dag_{-\vk+\vQ_p}|F_N\ran\equiv \Omega |i'\ran\,,\label{Udagphc_i}
\ee
where $\Omega$ is the vacuum Rabi coupling and $a^\dag_\vk$ ($ b^\dag_\vk$)  creates a free conduction electron (valence hole) with momentum $\vk$. We already see  that Pauli blocking forbids the photocreated pair to form a standard exciton because the  $|\vk|<k_F$ states are occupied,  the low-$\vk$ states  being actually responsible for most of the exciton binding.\

\begin{figure}[t]
\centering
  \includegraphics[trim=5.5cm 3.4cm 6.5cm 3cm,clip,width=2.4in] {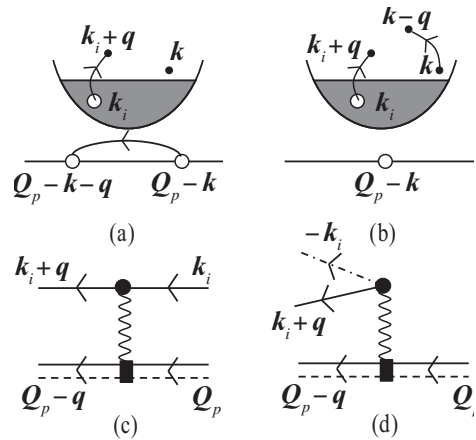}
\caption{\small The valence hole $(\vQ_p-\vk)$, as in (a), or the conduction electron $\vk$, as in (b),  can  scatter through Coulomb interaction with an electron $\vk_i$ inside the FS. Processes (a,b) lead to the scattering of a $\vQ_p$ exciton (solid-dashed double line) into a $(\vQ_p-\vq)$ state through the excitation of a FS electron $\vk_i$ to a $(\vk_i+\vq)$ state, as in  (c), or equivalently the creation of a FS electron-hole pair $(\vk_i+\vq,-\vk_i)$, as in  (d). Dot-dashed line represents FS hole. }
\label{fig:3}
\end{figure}

The $|i'\ran$ state in Eq.~(\ref{Udagphc_i}) is not an eigenstate of the semiconductor Hamiltonian because of the Coulomb interactions $V_{ee}$ between  conduction electrons and $V_{eh}$ between conduction electrons and the valence hole. The $V_{eh}$ interaction allows the photocreated electron to fly above the FS, as  in Fig.~\ref{fig:2}(b), or excites a FS electron-hole pair, as in Fig.~\ref{fig:3}(a). The $V_{ee}$ interaction can also excite a FS electron-hole pair, as  in Fig.~\ref{fig:3}(b). So, these two Coulomb interactions lead to\\
(i) processes in which the photocreated pair keeps its total momentum $\vQ_p$, and the FS stays unchanged, as  in Fig.~\ref{fig:2}(c), its role simply being to Pauli-block low-momentum states from participating in the ladder interactions that make an exciton,  in this way weakening its binding;\\
(ii) processes in which the photocreated pair changes its momentum from $\vQ_p$ to $(\vQ_p-\vq)$, while an electron $\vk_i$ from the FS is excited to  $(\vk_i+\vq)$ above the FS (see Fig.~\ref{fig:3}(c)), which is equivalent to saying that a FS electron-hole pair $(\vk_i+\vq,-\vk_i)$ is  created (see Fig.~\ref{fig:3}(d)).\\

Including more Coulomb processes would lead to the excitation of more FS electron-hole pairs. These multiple pair excitations  eventually change the step-like line shape of the excitation spectra near the onset of band-to-band transitions into a power-law divergence  known as Fermi edge singularity, significant for high doping only. For low and intermediate doping regimes as considered here,  states that involve zero or one FS electron-hole pair  shall provide  suitable trial states; then, the Fermi sea not only produces  a ``truncated" exciton, {\it i.e.}, an exciton without correlation from  low-$\vk$ electrons, but also reacts to the excitonic dipole through the excitation of one FS electron-hole pair.\

\begin{figure}[t]
\centering
 \includegraphics[trim=4cm 4.6cm 2.8cm 4.8cm,clip,width=3.2in] {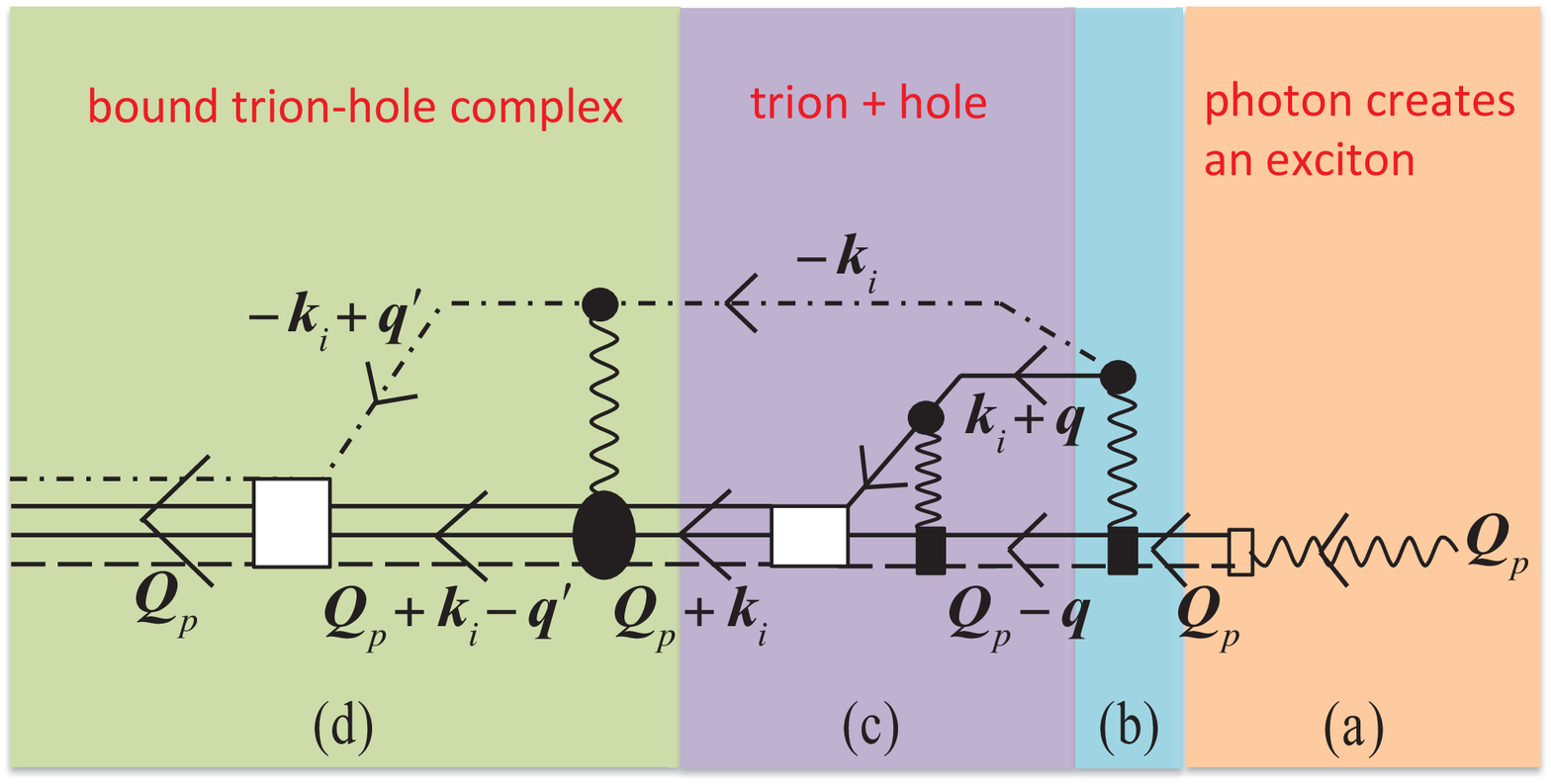}
\caption{\small The Coulomb interaction between the photocreated exciton $\vQ_p$ and a conduction electron $\vk_i$ in the FS produces a FS electron-hole pair $(\vk_i+\vq,-\vk_i)$, as in (b). A bound  trion can be formed through the repeated scatterings between the exciton and the FS electron flying above the FS, as in (c).  When the FS is sizable, the hole $(-\vk_i)$ left in the FS can swim inside the FS to  form a bound state with the trion, {\it i.e.}, a bound trion-hole complex, as in (d).}
\label{fig:4}
\end{figure}

Considering the excitation of just one pair already engages a very interesting new physics. Indeed, with four particles---one valence hole, one FS hole, and two electrons above the FS --- complex structures can emerge: the photocreated exciton can attract a conduction electron from the FS to form a bound trion if the two electrons have opposite spins (see Fig.~\ref{fig:4}(c)), as a result of repeated ladder-like interactions between the electron and the exciton, as shown in Fig.~\ref{fig:3}(c). Here also, the FS Pauli blocks low-momentum electron states from participating in the formation of a bound trion. So, compared to exciton, this sets an even stricter upper limit to the doping level  for such bound trion to exist.  The negatively-charged trion can further attract the FS hole to form a bound trion-hole complex (see Fig.~\ref{fig:4}(d)). For this attraction to produce a bound state, the  hole subspace must be large enough to allow repeated scattering, thus putting an even lower limit to the  Fermi-sea size. This leads to a rather narrow window for the trion-hole complex to exist. Note that,  the scenario in which the 4-particle system would form a  biexciton made of an exciton constructed with a valence hole and an exciton constructed with a FS hole, is not likely to occur because the latter pair does not form  an exciton due to the improper hole band structure. \

When the Fermi sea increases,  more electron  states are Pauli-blocked. In addition, this increase produces a stronger screening on the Coulomb interaction. For these two reasons, the bound trion  would end by dissociating into an electron and an exciton when the doped Fermi sea gets too large. However, at some stage prior to trion dissociation, the FS hole sets in with the yet-dissociated electron to dress the exciton into a more stable exciton-polaron, while the trion character stays in the excited states of the 4-particle complex.  As a result, when  the doping increases, the 4-particle system undergoes a transition from a trion-hole complex to an exciton-polaron. The main purpose of this work is to study when and how such a cross-over takes place, and how the excitation spectrum behaves through the cross-over.\

To determine this cross-over, we must derive the ground and low-lying excited states of the 4-particle system made of one valence hole, one FS hole and two electrons above the FS. This is best done through the Rayleigh-Ritz variational procedure. In this procedure, a set of restricted carrier bases  is used to find the Hamiltonian eigenstates, the precision of which increases with the number of states included in the basis. Throughout  this work,  the valence hole mass is  taken as infinite---a reasonable approximation since typical semiconductors have a valence hole mass  5 to 10 times larger than the conduction electron mass. This approximation enhances the physical effects associated with the formation of these 4-body complexes without qualitatively changing their physics.  Including a finite hole mass is possible but would require far heavier numerical work.\

So far, we have omitted the particle spin degrees of freedom. Yet, they are indispensable for the present problem  because trion bound states exist only when electrons have  opposite spins. In addition, while all particles interact via the same Coulomb interaction regardless of their spins, only same-spin fermions mutually suffer Pauli blocking. So, changing the polarization of the Fermi sea provides an interesting means to study resonance states.\


The Hamiltonian for the electron-hole  system we here consider   reads as
\be
H=H_0 + V_{ee}+V_{eh}\,.
\ee
For  infinitely heavy valence hole, the kinetic part reduces to $H_0=\sum_{\vk,s}\varepsilon_{\vk} a^\dagger_{\vk,s}a_{\vk,s}$, where  $a^\dagger_{\vk,s}$ creates a conduction electron with momentum $\vk$, spin $s=\pm1/2$ and energy $\varepsilon_{\vk}=\hbar^2 \vk^2/2m_e^*$ with $m_e^*$ being the conduction-electron effective mass. The  Coulomb interaction between conduction electrons is given by
\be
 V_{ee}=\frac 1 2 \sum_{\vq\not=0} v_{\vq} \sum_{\vk_1,s_1}\sum_{\vk_2,s_2} a^\dagger_{\vk_1+\vq,s_1}a^{\dagger}_{\vk_2-\vq,s_2}a_{\vk_2,s_2} a_{\vk_1,s_1}\, ,
\ee
while the electron-valence hole part is given by
\be
 V_{eh}= -\sum_{\vq\not=0} v_{\vq} \sum_{\vk_1,s_1}\sum_{\vk_2,s_2} a^{\dagger}_{\vk_1+\vq,s_1}b^{\dagger}_{\vk_2-\vq,s_2} b_{\vk_2,s_2}a_{\vk_1,s_1}\, ,
\ee
where $b^{\dagger}_{\vk,s}$ creates a $\vk$ valence hole with spin $s$.
The screened  Coulomb scattering  is taken as
\be
v_{\vq} ={\bar  v_{\vq} }/{\kappa_q}\,,\label{vq_kappa}
\ee
where
\be
 \bar v_{\vq} =\frac{2\pi e^2}{A \epsilon_{sc} q(1+qr_0)} \label{vqQW}
 \ee
 is the bare Coulomb potential in undoped 2D (when $r_0=0$) or quasi-2D (when  $r_0$ is finite)  semiconductors.
$A$  is the  2D sample  area and  $\epsilon_{sc}$ is the semiconductor static dielectric constant. The function $\kappa_q$  accounts for the Thomas-Fermi screening due to doped electrons. Its dependence on the FS electron density, through the Fermi wave vector $k_F$, is given by
$ \kappa_q=1+s_q/q(1+q r_0)$   with  $s_q=n_p/a_X$  for $q \le 2k_F $ and $s_q=n_p\big(1-\sqrt{1-(2k_F/q)^2}\big)/a_X $  for $q > 2 k_F$, with $n_p=1$ for fully polarized FS and $n_p=2$ for unpolarized  FS [see Ref.~\onlinecite{Stern}]. It has been shown that Eq.~(\ref{vqQW}) can suitably describe the Coulomb interaction in quasi-2D semiconductors such as III-V and II-VI QWs\cite{q2D,Sanders}. For an electron interacting with the heavy hole in GaAs/GaAs/Al$_x$Ga$_{1-x}$As QWs, the parameter $r_0$ varies from 2nm to 3nm for QWs with well width from 5nm to 8nm and $x=0.25$. Here we consider the case $r_0=0.3a_X$ with $a_X\simeq$10nm for GaAs and 3.3nm for ZnSe$^4$. The quasi-2D Coulomb potential used here is similar to the expression
$ \bar v_{\vq} =4 e^2/A \epsilon_{sc} q \tan^{-1}(q_L/q)$ suggested in Ref.~\onlinecite{Suris}, where $q_L$ is a parameter depending on the quantum-well width.
Note that in real space the quasi-2D Coulomb potential takes the analytic form\cite{q2D,Sanders}
\be \bar v(r) = \frac {Z(r/r_0)e^2} {\epsilon_{sc} r} \ee
with $Z(x)= \frac {\pi x} 2 \big({\bf H}_0(x)-N_0(x)\big)$, where ${\bf H}_0(x)$ and $N_0(x)$ denote the Struve and Neuman functions of zeroth order, respectively. As $r_0$ goes to zero, $Z(r/r_0)$ goes to 1.  As pointed out in Ref.~\onlinecite{Strinati}, the static screening is a valid approximation when the exciton binding is a small fraction of the band gap, as in III-V and II-VI semiconductor QWs. For 2D materials such as TMDs, the dynamic screening may have to be taken into account.  Our main purpose here is to present a detailed study of the exciton-trion-hole cross-over behavior for 2D and quasi-2D systems where static screening is a good approximation. We will not here delve into dynamic screening and band structure effects (including valley degrees of freedom and the spin splitting)\cite{Wang} which are important for TMDs.

\section{Basis for the Rayleigh-Ritz variational procedure\label{sec3}}
For a valence hole mass taken as infinite, the centers of mass of the exciton and the related complexes are fixed  on the valence hole. Their energies and wave functions are  associated with the relative motions of the light-mass conduction particles located at $\vr_i$ from the  valence hole. The set of basis functions to represent  conduction particles, with 2D polar coordinates $\vr=(r,\varphi_r)$, can be taken as
\be
\langle \vr|\phi_{n,m}\rangle=  e^{im\varphi_r} f_{n,m}(r)\, ,\label{FT:phiam}
\ee
where  $m=(0,\pm 1,\pm 2,\cdots)$ labels the angular part and $n=(0,\pm1,\pm2,\cdots)$ labels the radial part. By noting that the conduction particles bound to the valence hole have a node at the origin except when $m=0$, we take the radial part of the basis functions as
\be
f_{n,m}(r)= r^{(1-\delta_{m,0})} e^{-\alpha_n r}\, ,
\ee
the radial extension being controlled by the parameter $\alpha_n$ that we take as $\alpha_0 g^{n}$. A few $(m,n)$'s are sufficient to get fast convergence, for $\alpha_0$ and $g$  properly chosen  to  cover the physical range of interest. We find that  $g$, which corresponds to the ratio of two consecutive extensions, is better taken between $1.5$ and  $2$; the overlaps between the basis functions are then small enough to avoid numerical instability in the  variational calculation. In the present work, we have taken $g=2$ (1.8) when 8 (10) radial basis functions are used. Choosing a smaller $g$ would lower the ground-state energy slightly (e.g. by $10^{-4} R_X$ where $R_X$ is the 3D exciton ground-state energy), but the numerical computation will suffer larger round-off error when a large number of basis functions are needed. We wish to note that the basis functions are orthogonal with respect to the angular index but \textit{not} with respect to the radial index, that is, $\langle \phi_{n,m}|\phi_{n',m'}\rangle =0$ for $m\ne m'$ but not for $n\neq n'$.\

To study the effect of Pauli blocking by FS electrons in an easy way, we turn from $\lan \vr|\phi_{n,m}\ran$ to $\lan\vk|\phi_{n,m}\ran$. For $\vk=(k,\varphi_k)$, this 2D Fourier transform follows from
  \be
\lan \vk | \phi_{n,m} \ran =\int_{A} d^2r \lan \vk|\vr\ran \lan \vr| \phi_{n,m} \ran\equiv \frac 1 {\sqrt{A}} e^{im\varphi_k} F_{n,m}(k)\,,
\ee
with $\lan\vk |\vr\ran = e^{-i\vk \cdot \vr}/\sqrt{A}$. For a sample  disc with  area $A=\pi R^2$, the radial part then reads
 \be
 F_{n,m}(k)=\int_0^R \!\! r dr f_{n,m}(r) \!\int_0^{2\pi}\!\! d\varphi~  e^{-im\varphi}e^{-ikr\cos \varphi }\, .\label{Fm1}
\ee
The integral over $\varphi=\varphi_r-\varphi_k$ gives $(-i)^m 2 \pi J_m(kr) $, where
$J_m(kr)$ is the Bessel function of order $m$. As the remaining integrand is bounded by $f_{n,m}(r)$, the  $r$ upper boundary can be extended to infinity. Then, the result simply follows from first or second derivative over ($-\alpha_n$) of the same integral calculated without the $r^{2-\delta_{m,0}}$ factor. This leads to
\be
F_{n,m}(k) {=} {2\pi} ({-}i)^m \!\left(\!{-}\frac{\partial}{\partial \alpha_n}\right)^{2{-}\delta_{m,0}}\! \left[\frac{(\sqrt{\alpha_n^2+k^2}{-}\alpha_n)^{|m|}}{k^{|m|}\sqrt{\alpha_n^2{+}k^2}}\right]\,. \label{Fm}
\ee

\section{Exciton with a Frozen Fermi sea\label{sec4}}

Let us first consider how the photocreated exciton, made of a conduction electron with spin $(-1/2)$ and a valence hole with angular momentum $(3/2)$ is affected by the occupied electron states in a Fermi sea,  $|F_{N_+,N_-}\ran$, having $N_+$ electrons with spin $(1/2)$ and $N_-$ electrons with spin $(-1/2)$. In this section, we shall focus on the effect of Pauli blocking induced by the FS on this exciton. So, for the moment we forget  electron-hole pair excitations from the Fermi sea due to Coulomb interaction with the exciton, that is, we take the Fermi sea as frozen.\

We obtain the exciton eigenstates $i$ by solving the Schr\"{o}dinger equation
 \be
 (H-E_i^{(e)})|\Psi^{(e)}_i\ran=0\,.\label{12}
  \ee
 Using the basis functions that we have previously introduced  for conduction electrons, we expand  exciton states $|\Psi^{(e)}_i\ran$ with angular momentum $m_i$
 \be
|\Psi^{(e)}_i\ran =\sum_n x^{(i)}_{n,m_i} |n,m_i\ran  \label{X0}
 \ee
 on conduction electron-valence hole pair states taken as
 \be
 |n,m\ran = \sum_{\vk}\lan \vk | \phi_{n,m} \ran
 a^{\dagger}_{\vk,-\frac 1 2}b^{\dagger}_{\vQ'_p,\frac{3}{2}}|F_{N_+,N_-}\ran \,, \label{X1}
 \ee
where $\vQ'_p+\vk $ is equal to the absorbed photon momentum $\vQ_p$ due to momentum conservation kept by  Coulomb processes. Note that Pauli blocking with the Fermi sea imposes $|\vk|$ in the above sum to be larger than the Fermi momentum $k_{F_-}$ of the $N_-$ electrons.\

\begin{figure}[t]
\centering
\subfigure[]{\label{fig:5a} \includegraphics[trim=0cm 0cm 1cm 0.5cm,clip,width=3in] {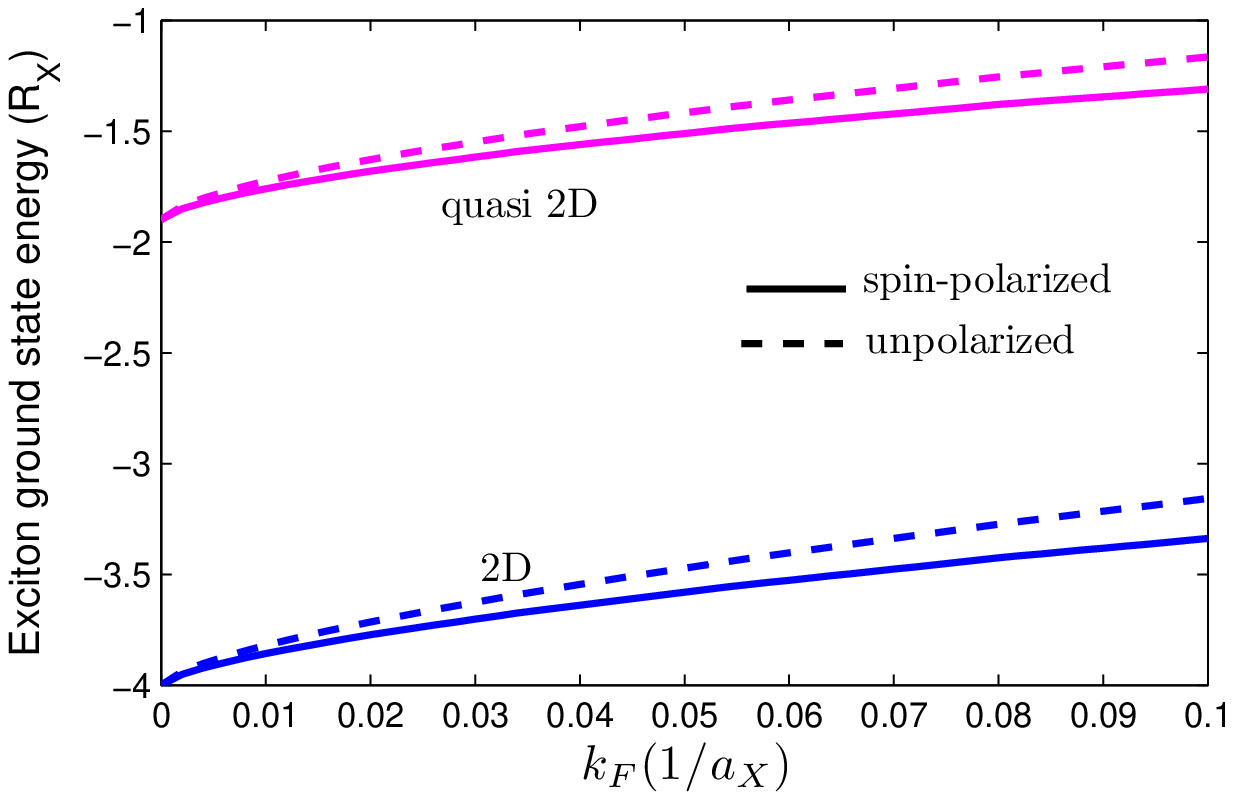}}
\subfigure[]{\label{fig:5b} \includegraphics[trim=0cm 0cm 0.5cm 0.5cm,clip,width=3in] {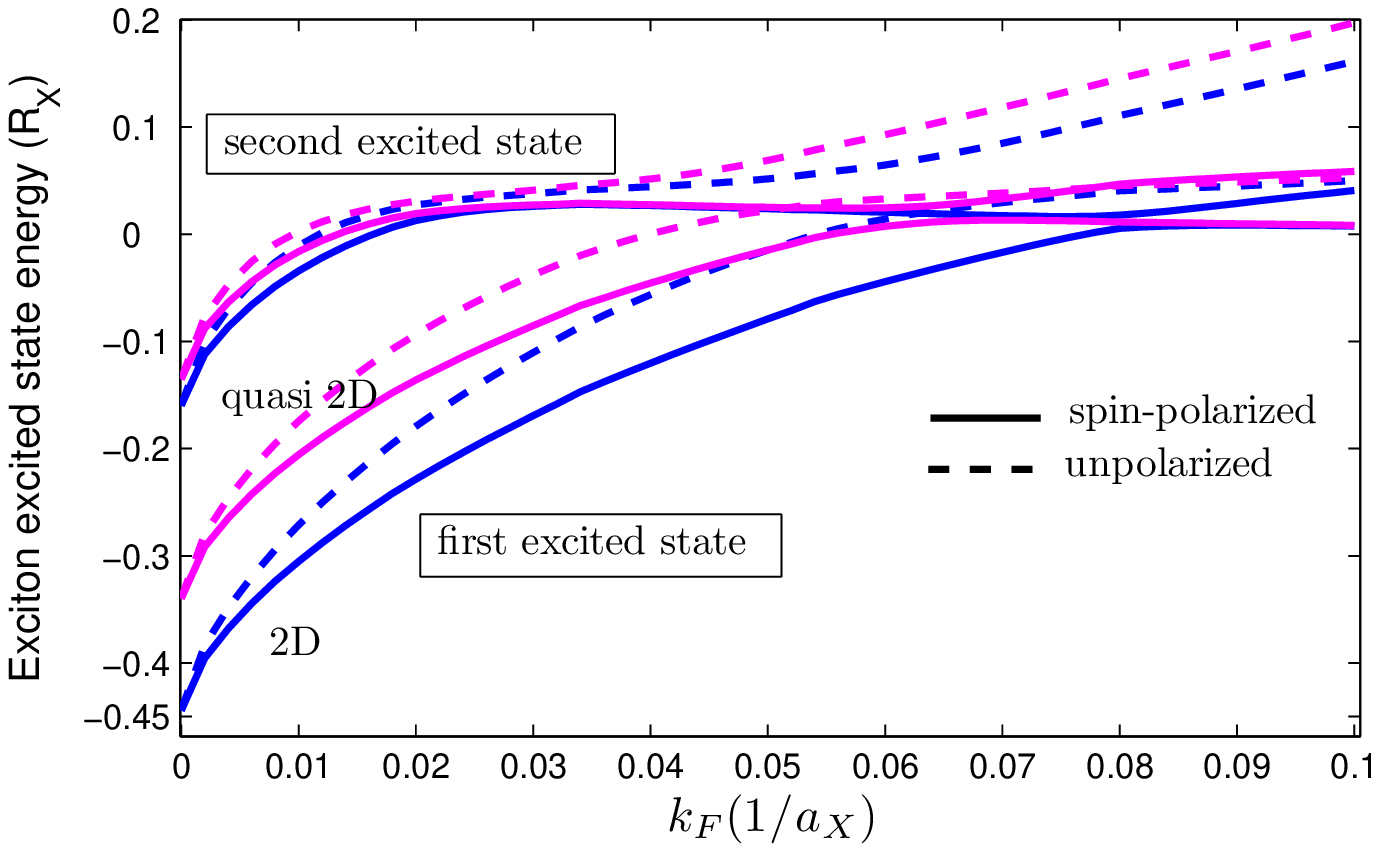}}
\caption{\small Energies of the exciton ground state (a) and first two excited states (b) with angular momentum $m=0$ as functions of $k_F$ for 2D (blue) and quasi-2D (pink) QW. Solid (dashed) curves are for spin-polarized (unpolarized) Fermi sea.}
\label{fig:5}
\end{figure}

We project the above $|\Psi^{(e)}_i\ran$ over $\lan n',m_i|$. The projection over states with $m'_i\neq m_i$ readily gives zero due to the orthogonality of the $| \phi_{n,m} \ran $ states. This leads to a set of coupled equations
  for the $x^{(i)}_{n,m_i}$ coefficients,
  \be
   \sum_n \lan n',m_i|H-E_i^{(e)} |n,m_i\ran x^{(i)}_{n,m_i}=0\,, \label{X}
    \ee
which depend on the number of $|n,m_i\ran$ basis states taken in the $|\Psi^{(e)}_i\ran$ expansion.
By numerically solving the resulting equation,  we can compute the eigenstates and eigenvalues $E_i^{(e)}={\cal E}_i^{(e)}+ \mathcal{E}_{N_+,N_-}$ for excitons with angular momentum $m_i$  in the presence of the $|F_{N_+,N_-}\ran$ Fermi sea having energy $\mathcal{E}_{N_+,N_-}$. The matrix elements needed to perform this calculation are given in \ref{app:sec1}.

$\bullet$ \textbf{In the absence of Fermi sea}, that is, for $N_+=N_-=0$, the  2D exciton energies are analytically known\cite{2DX} as ${\cal E}^{(e)}_n=-(n-1/2)^{-2}R_X$ labeled with the principal quantum number $n$, where $R_X$ is the 3D exciton ground-state energy.  The Rayleigh-Ritz variational procedure we here use  gives this analytical exciton  ground-state energy by just taking {\it one} radial basis function with  $\alpha_0=2a_X^{-1}$.  By taking 10 radial basis functions with $n=(0, 1,  2,\cdots 9)$ and $\alpha_0=0.125a_X^{-1}$, and  5 angular basis functions with $m=(0,\pm 1, \pm 2)$, we also obtain the energies of the $2s, 2p$, and $3d$ excited states in agreement with the analytical 2D exciton values within an error less than $2 \times 10^{-4}R_X$. For  quasi-2D QW, the exciton binding is approximately one half of that in the 2D case, a result consistent with the previous finding\cite{Sanders} for GaAs/Al$_x$Ga$_{1-x}$As QW with width around 10nm and $x=0.25$.

\begin{figure}[t]
\centering
\subfigure[]{\label{fig:6a} \includegraphics[trim=1cm 1cm 1cm 14cm,clip,width=3.in]  {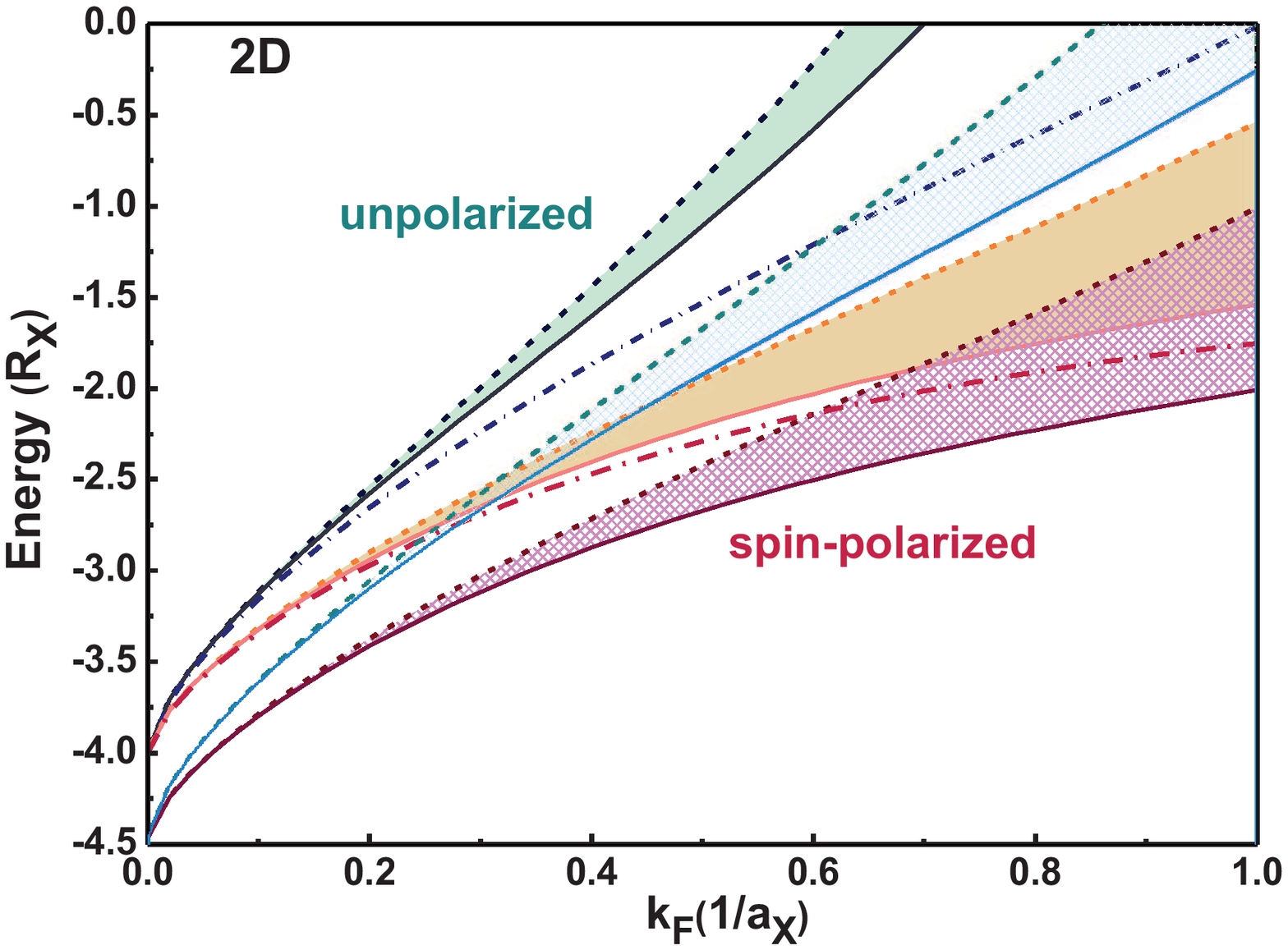}}
\subfigure[]{\label{fig:6b} \includegraphics[trim=0cm 0cm 0cm 0cm,clip,width=3in]  {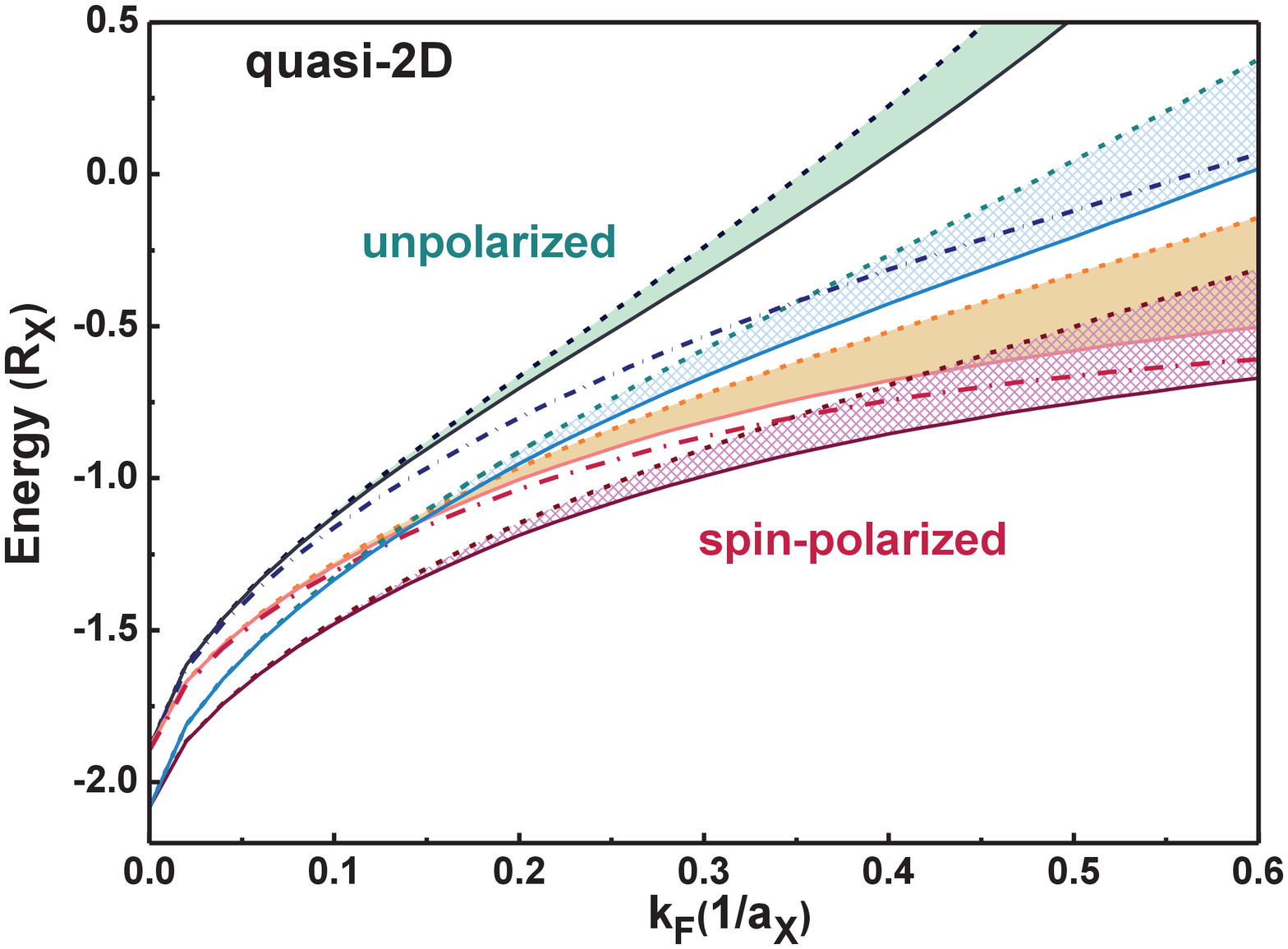}}

 \vspace{-0.1cm}
\caption{\small (a) 2D and  (b) quasi-2D QWs. In both cases, the energies of ground-state exciton (red and dark-blue dash-dotted curves), ground-state trion (brown and green dashed curves), trion minus $E_F$ (brown and light-blue solid curves), trion first excited state (orange and black dashed curves), and  trion first excited state  minus $E_F$ (orange and black solid  curves) are shown. The lower (upper) group of curves is for spin-polarized (unpolarized) FS. The shaded areas indicate a continuum energy range of all possible FS-hole states associated with a given trion state when the interaction between the trion and FS hole states is absent. The upper and lower groups of curves correspond to unpolarized and spin-polarized FS, respectively. }
\label{fig:6}
\end{figure}

$\bullet$ \textbf{In the presence of a fully polarized Fermi sea}, that is, for $N_+=N$ and $N_-=0$, the photocreated electron does not feel Pauli blocking from the Fermi sea. Still, a screening effect produced by the conduction electrons goes to weaken the strength of the Coulomb potential, and consequently reduces the binding energies of all bound states.  Figure \ref{fig:5} shows a significant reduction of the binding energies of the exciton ground and first two excited states when the doping density increases, as obtained by using the Thomas-Fermi screening in Eq.~(\ref{vq_kappa}).  We find that the first (second) excited state  becomes unbound  when $k_Fa_X$ exceeds  $0.078 ~(0.016)$ for 2D QW. For quasi-2D QW, they become unbound when $k_Fa_X$ exceeds $0.056$ and $0.014$, respectively. By contrast, the ground states for 2D and quasi-2D remain bound for all values of $k_F$ considered (see the  red dash-dotted curves in Fig.~\ref{fig:6}).\

$\bullet$ \textbf{In the presence of an unpolarized Fermi sea}, that is, for $N_+=N_-=N/2$, not only the Coulomb potential is screened, but also the  $\vk$ sum in Eq.~(\ref{X1}) is restricted to  $|\vk|$ larger than the Fermi  momentum $k_F$. We performed the calculation using basis functions $F_{n,m}(k)$ given in Eq.~(\ref{Fm}) for $n=(0,1,2,...,9)$ with $k>k_F$ and $\alpha_0=k_F+0.125a_X^{-1}$. The minimum value of $\alpha_n$ is chosen to be slightly larger than $k_F$, since any basis function with $\alpha$ smaller than $k_F$ will have too strong overlap with the leading basis function due to the cut-off at $k=k_F$. When $k_F=0$, this basis set coincides with the one we have used  in the absence of Fermi sea. The computed energies of the three lowest-lying states as  functions of $k_F$ are  shown in Fig.~\ref{fig:5}. The  first (second) excited state  becomes unbound  when $k_Fa_X$ exceeds  $0.054 ~(0.012)$ and $0.04 ~(0.01)$ for 2D and quasi-2D QWs, while the ground states remain bound for all values of $k_Fa_X$  considered (see the  dark-blue dash-dotted curves in Fig.~\ref{fig:6}).

\section{Trion with a frozen Fermi sea\label{sec5}}

We now consider the possibility of forming a trion by the Coulomb interaction between a conduction electron and the excitonic  pair created by a $\sigma_+$ photon, in the presence of a frozen Fermi sea $|F_{N_+,N_-}\ran$. By again taking the Fermi sea as frozen, we continue our  focus on the effect of Pauli blocking on the 3-particle complex. First, we note that in order to possibly form a bound trion, the two conduction electrons of this complex must have opposite spins. So, the system we consider actually has $(N_++1)$ up-spin electrons and $(N_-+1)$ down-spin electrons. For $(N_+=0,N_-=0)$, the system reduces to the conventional $X^-$ trion.\

 To solve the Schr\"{o}dinger equation
 \be
 (H-E^{(ee)}_i)|\Psi^{(ee)}_i\ran=0 \label{16}
 \ee
  for this 3-particle complex, we again expand $|\Psi^{(ee)}_i\ran$ on the basis functions for conduction electrons introduced in Sec.~\ref{sec3}. By noting that the states possibly seen in photo-absorption have a total angular momentum equal to zero, this expansion can be reduced to
\be
|\Psi^{(ee)}_i\ran=\sum_m \sum_{n_1,n_2}  t^{(i)}_{n_1,n_2,m}|n_1,n_2,m\ran \, ,\label{Tr}
\ee
the 3-particle basis  being taken as
\bea
|n_1,n_2,m\ran &=& \sum_{\vk_1,\vk_2}\phi_{n_1,m}(\vk_1)  \phi_{n_2,-m}(\vk_2) \nn \\
&\times& a^{\dagger}_{\vk_1,-\frac 1 2}a^{\dagger}_{\vk_2,\frac 1 2}b^\dag_{\vQ'_p,\frac{3}{2}}|F_{N_+,N_-}\ran\,,\label{states:n1n2m}
\eea
with $\vQ'_p+\vk_1+\vk_2=\vQ_p+\vK$ where $\vK$ is the initial momentum of the spin $(1/2)$ electron, due to momentum conservation in Coulomb processes. Note that Pauli blocking forces the  above sums over $\vk_1$ and $\vk_2$  to start with values  larger than the Fermi momenta of the $(N_+,N_-)$ electrons, respectively.

We use this $|\Psi^{(ee)}_i\ran$  in the Schr\"{o}dinger equation (\ref{16}) for the 3-particle complex and we project it over $\lan n'_1,n'_2,m'|$. This leads to a set of linear equations
 for the $t^{(i)}_{n_1,n_2,m}$ coefficients
\be
\sum_{n_1,n_2,m} \lan n'_1,n_2',m'|H-E^{(ee)}_i|n_1,n_2,m\ran t^{(i)}_{n_1,n_2,m} =0\,.\label{HT}
\ee
We  numerically solve  the above eigenvalue problem to obtain the eigenvalues $E_i^{(ee)}={\cal E}_i^{(ee)}+\mathcal{E}_{N_+,N_-}$. Explicit expressions of the matrix elements needed to perform this calculation are given in \ref{app:sec2}.\

We wish to note that, for infinite valence-hole mass, the prefactors in Eq.~(\ref{states:n1n2m}) do not depend on $\vQ'_p$, or equivalently, on $\vK$; so, the obtained eigenstates do not depend on the initial momentum of the spin-$(1/2)$ electron added to the photocreated pair. This comment will be of importance in the next section.\

$\bullet$ \textbf{In the absence of Fermi sea}, the problem reduces to a conventional trion in 2D  or quasi-2D systems. The two electrons with spin $(1/2)$ and $(-1/2)$  do not suffer  Pauli blocking nor  Coulomb screening. For  2D, the computed trion  ground-state energy ${\cal E}_g^{(ee)}$ is equal to  $-4.47 R_X$ by taking $(n_1,n_2)=(0,1,\cdots,7)$, $\alpha_0 =0.125a_X^{-1}$, and $m=(0,\pm 1, \pm 2)$. This gives a binding energy of  trion with respect to  exciton  equal to $0.47 R_X$, in remarkable  agreement with the best variational results\cite{Stebe1989,Thilagam1997,Sergeev2001}. When the Fermi sea is present, the same $(n,m)$'s and  $\alpha_0$ replaced by $k_F+0.125a_X^{-1}$ show fast convergence for the low-lying states of interest.

$\bullet$ \textbf{In the presence of a fully polarized Fermi sea}, that is, for $N_+=N$ and $N_-=0$,  the photocreated exciton does not suffer Pauli blocking from the FS, but the spin-$(1/2)$ electron does.  The computed trion ground-state energies ${\cal E}_g^{(ee)}$  for 2D  or quasi-2D systems as functions of  $k_F$ are shown by the red  dashed curves in Fig.~\ref{fig:6}.

$\bullet$ \textbf{For an unpolarized Fermi sea},  $N_+=N_-=N/2$, the trion is formed from a linear combination of $(\vk_1,\vk_2)$ electrons which are both outside their respective Fermi seas. The computed trion ground-state energies for 2D  or quasi-2D systems as functions of $k_F$ are  shown by the dark-blue dashed curves in Fig.~\ref{fig:6}.

In the presence of a Fermi sea, we expect that the effects of Pauli blocking and Coulomb screening grow with increasing doping, to ultimately cause the trion to dissociate into an exciton plus a spin-$(1/2)$ electron sitting on top of the FS ({\it i.e.} with energy $E_F$). So, to compare with the exciton ground-state energy in the presence of a Frozen FS as indicated by the dark-blue (red) dash-dotted curve for unpolarized (spin-polarized) FS, we should subtract $E_F$ from the trion ground-state energy, shown as the green (brown)  dashed curve in Fig.~\ref{fig:6}, and the resulting energy is shown as the light-blue (brown)  solid curve. For both spin-polarized and unpolarized Fermi seas, the trion remains bound over the entire doping range considered here (with $k_Fa_X \le 1$). As we continue to increase $k_F$, the trion will dissociate and the trion ground-state energy minus $E_F$ (light-blue and brown  solid curves)  will merge with the exciton ground-state energy (dark-blue and red dash-dotted curves) in the variational calculation.\

 To help understand the photo-absorption spectrum of this complex system, we also show  in Fig.~\ref{fig:6} the energy of the trion first excited state (black and orange dashed curves) and the same energy minus $E_F$ (black and orange solid curves). The trion first excited state merges with the exciton ground state at $k_F=0$, indicating no bound excited state for the negatively-charged trion in the large valence-hole mass limit. We see that for spin-polarized FS, the trion first-excited-state energy minus $E_F$ (orange solid curve) crosses the trion ground-state energy (brown dashed curve) near $k_Fa_X = $ 0.68~(0.42) for  2D (quasi-2D) system. No such crossing occurs for unpolarized FS. As will be shown later, this crossing will lead to  absorption spectra distinctly different  from those for unpolarized FS.


\section{Trion-hole complex \label{sec6}}

In this  section, we consider the possibility that the photocreated exciton can excite one electron out of the Fermi sea, leaving a FS hole (see Fig.~\ref{fig:3}). This possibility can be included by considering states like
\be
a^{\dagger}_{\vk_2,s}a^{\dagger}_{\vk_1,-\frac 1 2}b^\dag_{\vQ'_p,\frac{3}{2}}a_{\vk_i,s }|F_{N_+,N_-}\ran\, , \label{21}
\ee
its total momentum $\vk_1+\vk_2+\vQ'_p-\vk_i$ being equal to the momentum $\vQ_p$ of the photocreated exciton and $s=\pm  1 /2$. The system we consider now has $N_-+1$ spin-$(-1/2)$ electrons but $N_+$ spin-$(1/2)$ electrons only, instead of $N_+ +1$ as in the previous section. First, we note that the above state must have the $(\vk_i, s)$  electron  inside the $N_s$ Fermi sea. When $\vk_2=\vk_i$, the above state (\ref{21}) then contains a photocreated electron-hole pair, $a^{\dagger}_{\vk_1,- 1/ 2}b^\dag_{\vQ'_p,3/2}$, and a full Fermi sea  $|F_{N_+,N_-}\ran$, while for $\vk_2\neq\vk_i$, this state  has a $\vk_2$ electron above the $N_s$ Fermi sea and a $(-\vk_i)$ hole inside. States having more than one FS electron-hole pair are here neglected because we are only interested in low to intermediate doping regime.\


We look for the system eigenstates
\be
(H-E^{(eeh)}_i)|\Psi^{(eeh)}_i\ran =0\label{Schrod:dTr}
\ee
that have a zero total angular momentum.  In general, we can expand the eigenstates of the system in terms of two sets of basis states.
The first set contains exciton states (with  spin-$(-1/2)$ electron) in the presence of a rigid FS, while the second set contains spin-$(-1/2)$ single electron states accompanied by all possible single electron-hole pair excitations from the FS, $|F_{N_+,N_-}\ran$. We write
\bea
&&|\Psi^{(eeh)}_i\ran = \sum_j  z_{X,j}^{(i)}  |\Psi^{(e)}_{j}\ran \label{dTr} \\
&+& \sum_{n_1,n_2,\ell,m_1,m_3,s} z^{(i)}_{n_1,n_2,\ell,m_1,m_3,s}|n_1,n_2,\ell,m_1,m_3,s\ran \nn ,
\eea
where
\bea && |n_1,n_2,\ell,m_1,m_3,s\ran =\sum_{\vk_1,\vk_2,\vq}\phi_{n_1,m_1}(\vk_1)\phi_{n_2,-m_1-m_3}(\vk_2) \nn \\ && \times G_{\ell,m_3}(\vq)a^{\dagger}_{\vk_2,s}a^{\dagger}_{\vk_1,-\frac 1 2}b^\dag_{\vQ'_p,\frac{3}{2}}a_{\vq,s}|F_{N_+,N_-}\ran\,. \label{m3} \eea
In order to identify the trion character contained in the low-lying eigenstates, we transform the basis set with $m_3=0$ and $s=1/2$ into the trion-hole basis set, which contains products of trion eigenstates  found in the previous section and FS hole states with zero angular momentum. Namely, we replace the part
$\sum_{n_1,n_2,\ell,m_1} z^{(i)}_{n_1,n_2,\ell,m_1,0,\frac{1}{2}}|n_1,n_2,\ell,m_1,0,\frac{1}{2}\ran$ in the above expansion by
\be \sum_j \sum_{|\vq|<k_{F_+}} z^{(i)}_{T,j}(\vq)   a_{\vq,\frac 1 2}|\Psi^{(ee)}_j\ran \,,  \label{dTr1}
\ee
where \be
z^{(i)}_{T,j}(\vq)=\sum_{\ell} z^{(i)}_{T,j,\ell}  \,G_{\ell,0}(\vq) \,.
\ee
The remaining basis states correspond to single electron-hole  excitations from the spin-$(1/2)$ FS with $m_3\ne 0$ or from the spin-$(-1/2)$ FS with all possible $m_3$, if $N_- \neq 0$.\

\begin{figure}[t]
\centering
 \includegraphics[trim=2.8cm 4.8cm 3cm 4.8cm,clip,width=3in] {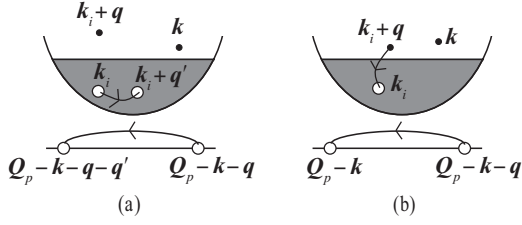}
 \vspace{-1cm}
\caption{\small We start from the configuration of Fig.~\ref{fig:3}(a), with  $(\vk,-1/2)$ and  $(\vk_i+\vq,1/2)$ electrons above the FS, a $(-\vk_i,-1/2)$ hole inside the FS, and a $(\vQ_p-\vk-\vq,3/2)$ valence hole. (a) The FS hole $\vk_i$ scatters into a $(\vk_i+\vq')$ state inside the FS through Coulomb interaction with the valence hole. This valence hole can have similar Coulomb interactions with the $\vk$ and $(\vk_i+\vq)$ electrons outside the FS (not shown). (b) Through Coulomb interaction with the valence hole,  the FS electron-hole pair $(\vk_i+\vq,-\vk_i)$  can also recombine. This leaves a conduction electron-valence hole pair $(-1/2,3/2)$ with momentum $(\vk,\vQ_p-\vk)$ in the presence of the full Fermi sea, just after the photocreation of a $\vQ_p$ exciton.   }
\label{fig:7}
\end{figure}

In the above equation, a finite set of basis functions $G_{\ell,m_3}(\vq)$ is further needed to describe  Coulomb scatterings between the FS hole $\vq$ in the  Fermi sea and the three particles of the  trion [See Fig.~\ref{fig:7}(a)]. $G_{\ell,m}(\vq)$ is taken as a set of orthogonal basis functions
\be G_{\ell,m}(\vq)=(-i)^m e^{im\varphi_\vq} \sqrt{\frac {2\pi(2-\delta_{\ell, 0})}{k_Fq}}~\cos\frac{\ell\pi q}{k_F}. \ee
These scatterings ultimately lead to a trion-hole complex. The trion-hole states are coupled to the exciton states $|\Psi^{(e)}_{j}\ran$ in  Eq.~(\ref{dTr}) in the eigenstates of the 4-particle complex by the Coulomb interaction, as shown in Fig.~\ref{fig:7}(b).

Since both the system eigenstates $|\Psi^{(eeh)}_i\ran$ and the trion states $|\Psi^{(ee)}_j\ran$ in Eq.~(\ref{dTr}) have zero total angular momentum,  the trion-hole envelope function $ z^{(i)}_{T,j}(\vq)$ must be $s$-like. The angular correlations between  trion and FS hole with $m_3\neq 0$ are included in the remaining terms of Eq.~(\ref{dTr}). We also wish to note that the scattering between  trion and FS hole induces a change to the trion momentum. However, when the valence hole mass is infinite, the trion states no longer depend on their center-of-mass momentum. This is why we can  use the previously  obtained trion states in Eq.~(\ref{dTr1}).\

We now project the Schr\"{o}dinger equation (\ref{Schrod:dTr}) onto the exciton states $|\Psi^{(e)}_{j}\ran$, the trion-hole states $a_{\vq',\frac 1 2}|\Psi^{(ee)}_{j}\ran$, and the remaining basis states of the form $|n_1,n_2,\ell,m_1,m_3, s \ran$. This leads to a set of coupled equations, from which we determine the $z^{(i)}_{n_1,n_2,\ell,m_1,m_3,s}$ coefficients that enter $|\Psi^{(eeh)}_i\ran$.  In numerically solving Eq.~(\ref{Schrod:dTr}), we have used 8 basis states with $n=(0,1,\cdots, 7)$ and $m=0$ for $|\Psi^{(e)}_j\ran$, twenty $G_{\ell,0}(\vq)$ functions with $\ell=(0,1,\cdots, 19$), multiplied by the lowest 64  trion eigenstates obtained from solving Eq.~(\ref{16}) in the basis defined in Eq.~(\ref{states:n1n2m}). Also included are the basis states
$|n_1,n_2,\ell,m_1,m_3, s \ran$ with $(n_1,n_2)=(0,1,\cdots, 7), \ell=(0,1,\cdots, 9)$, and $(m_1, m_2, m_3)=(0,\pm 1, \pm 2)$, while keeping $m_1+m_2+m_3=0$. For spin-polarized FS with $N_-=0$, only the additional basis states with $s=1/2$ and $m_3=(1,2)$ are included, while for unpolarized FS, we need to include  basis states with $s=-1/2$ and $m_3=(0,1,2)$.



The $|\Psi^{(eeh)}_i\ran$ state is trion-hole like when the squared amplitude $f_T^{(i)}=\sum_{|\vq| <k_F}|z^{(i)}_{T,g}(\vq)|^2$ containing the ground-state trion component is close to 1: The trion character then dominates. Here $i$ labels the eigenstates of the 4-particle complex. In the other limit, $f_X^{(i)}=|z_{X,g}^{(i)}|^2$ containing the ground-state exciton component  is much larger than  $f_T^{(i)}$, and  the $|\Psi^{(eeh)}_i\ran$ state is better seen as exciton-polaron like, that is, an exciton dressed by  FS electron-hole pairs.\

\begin{figure}[t]
\centering
\subfigure[]{\label{fig:8a} \includegraphics[trim=1cm 1cm 1cm 14cm,clip,width=3in]  {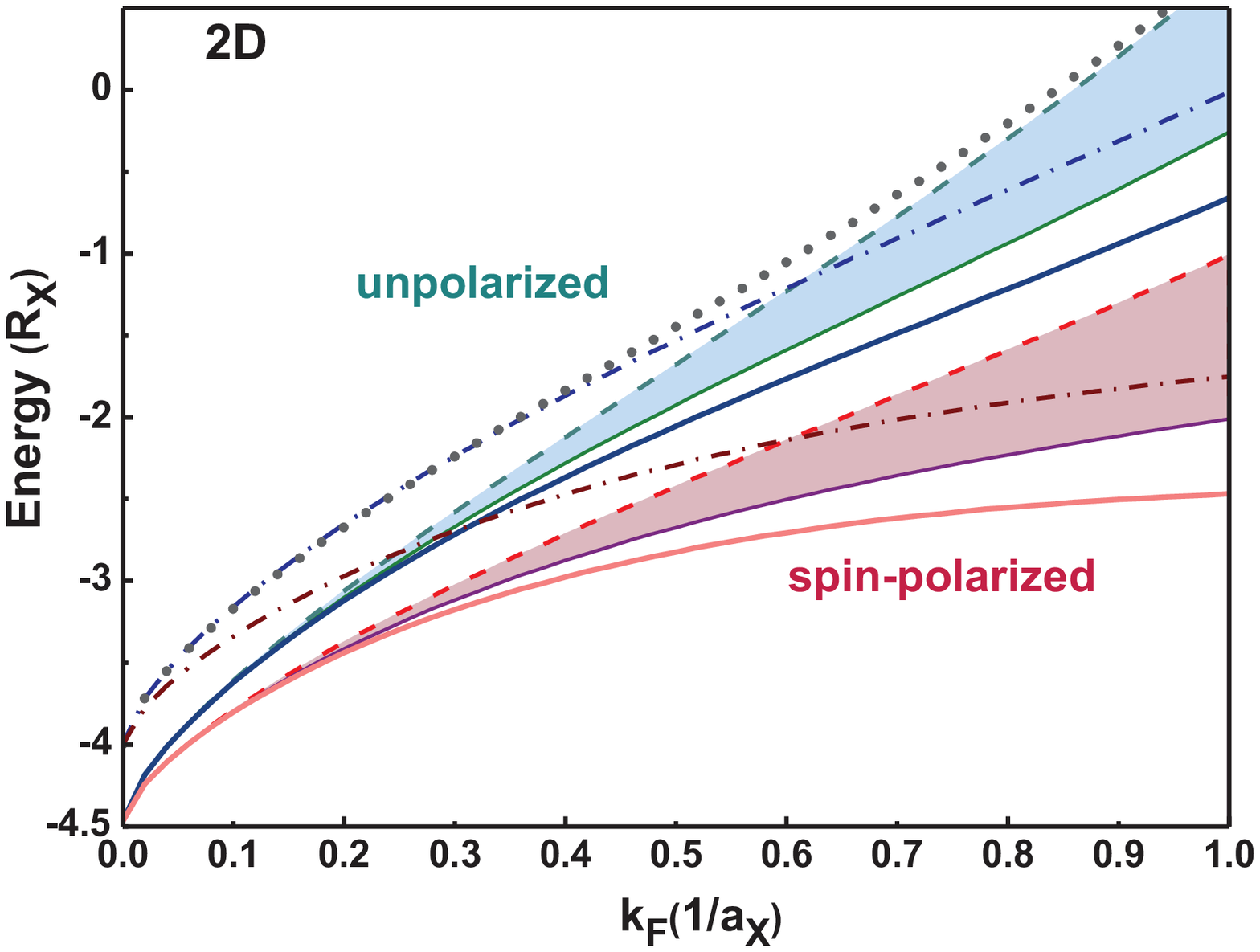}}
\subfigure[]{\label{fig:8b} \includegraphics[trim=1cm 1cm 1cm 14cm,clip,width=3in]  {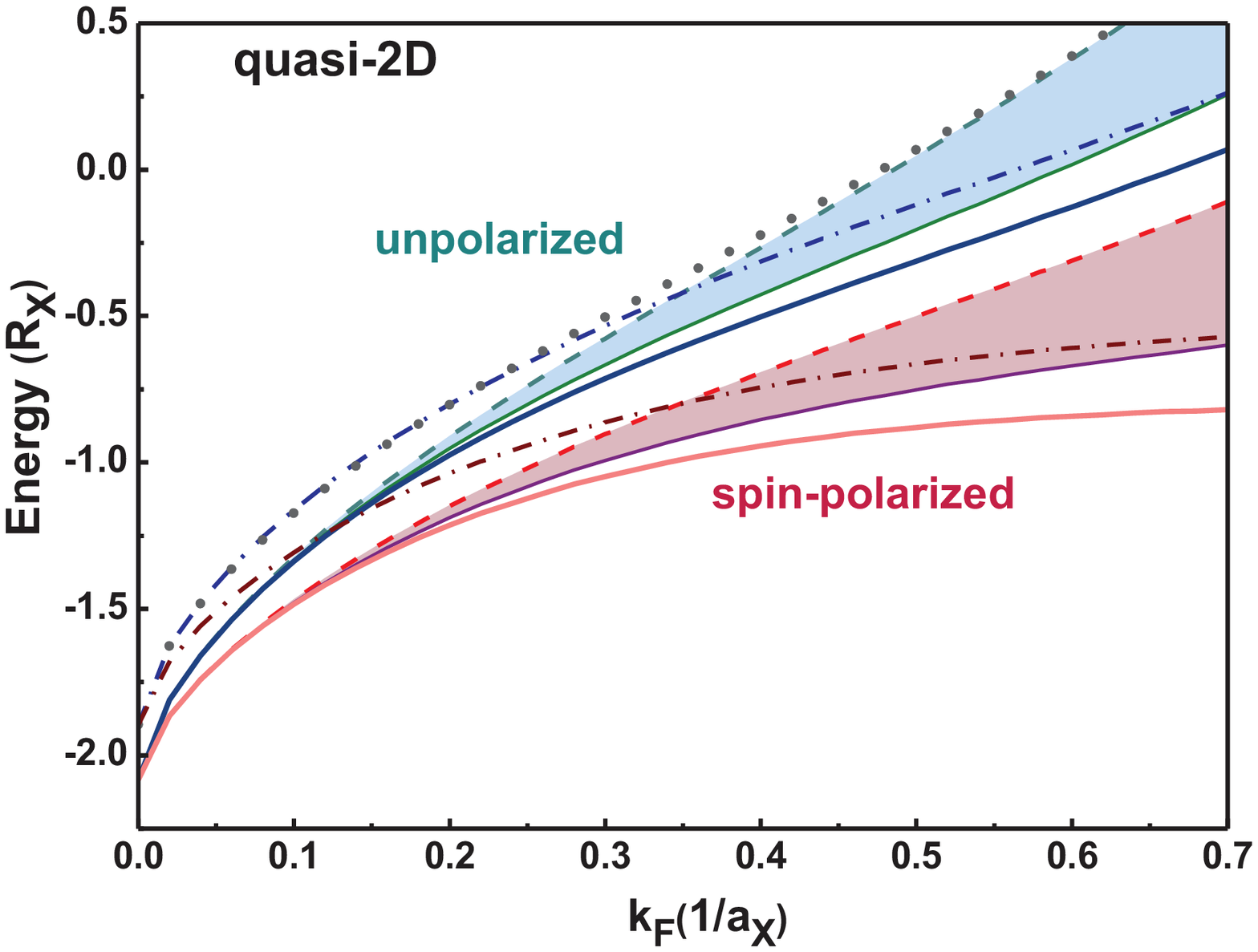}}
 \vspace{-0.1cm}
\caption{\small (a) 2D (b) quasi-2D QW. In both cases, the energies of ground-state exciton (brown and blue dash-dotted curves), ground-state trion (red and green dashed curves), trion minus $E_F$ (purple and green thin solid curves), and trion-hole complex evolving to exciton-polaron (orange and dark-blue thick solid curves) are shown. The lower (upper) group of curves is for spin-polarized (unpolarized) Fermi sea. The shaded area indicates a  continuum energy range of all possible trion-hole states when the interaction between the trion and FS hole states is absent. The solid  spheres delineate the energy positions of the upper branch of the coupled mode between trion-hole complex and exciton-polaron for unpolarized  FS, which correspond to the second peak in the absorption spectra in Figs.~\ref{fig:9}(a) and Fig.~\ref{fig:10}(a)}
\label{fig:8}
\end{figure}

 Figure \ref{fig:8} shows the ground-state energy of the 4-particle complex  as a function of $k_F$ as the orange (dark-blue) thick solid curve for spin-polarized (unpolarized) FS. The red (green) dashed curve represents the trion ground state plus a hole at the bottom of the spin-polarized (unpolarized) FS, while the purple (green) thin solid curve  represents the trion ground state plus a hole at the top of the FS, when the interaction between  trion and  FS hole is absent. So, the energy for the thin solid curve is  equal to the trion ground-state energy minus the Fermi energy, $E_F=\hbar^2 k_F^2/(2m_e^*)$.  Between the dashed and thin solid curves  is a continuum of states (indicated by shaded areas) that consist of the trion ground state and a FS hole having energy between 0 and $-E_F$. When the interaction between trion and  FS hole is taken into account, the energy of  the 4-particle complex ground state lowers to the thick solid curve. The fact that this curve is well separated from the continuum shows that the 4-particle complex forms a bound state. We note that  for both spin-polarized  and unpolarized Fermi seas, the dashed curve  crosses the dash-dotted curve (which indicates the energy of the exciton ground state with a Frozen FS)   at   $k_Fa_X \simeq 0.6 $  for  2D QW and at $k_Fa_X \simeq 0.34$ for  quasi-2D QW. Beyond the crossing point, the exciton ground-state level merges into the continuum of  trion-hole states (shaded areas) and  the two different species become strongly coupled. This also signals the advent of the cross-over from trion-hole complex to exciton-polaron.\

As discussed in the introduction, the character of this 4-particle ground state as a function of $k_F$ can be revealed by  the squared amplitude of the trion component, $f_T^{(0)}$  and the squared amplitude $f_X^{(0)}$ of the exciton component shown in Fig.~\ref{fig:1}, with  a cross-over  for spin-polarized (unpolarized) FS occurring at $k_Fa_X=0.81 ~(0.72)$ for 2D  and at $0.45 ~(0.4)$ for quasi-2D. For doping densities above the cross-over, the ground state of the 4-particle complex maintains a strong exciton component, as evidenced by its squared amplitude  shown in Fig.~\ref{fig:1}. The fact that this energy is lower than the exciton state by $\approx 0.7R_X~ (0.25R_X)$ for 2D (quasi-2D) indicates a binding of the exciton with a pair of FS electron and hole, which can be seen as exciton-polaron in the weak polaron limit.  The sum of $f_T^{(0)}$ and $f_X^{(0)}$ is very close to 1 at low doping but  becomes less than 1 at high doping; the deviation from 1 is attributed to the fraction contributed by the exciton excited states with amplitude $z^{(0)}_{X,j\neq g}$ and the trion excited states  with amplitude $z^{(0)}_{T,j\neq g}(\vq)$ in  the 4-particle ground state. Obviously, these contributions become more significant as the doping increases.
It is interesting to note that  for spin-polarized (unpolarized) FS, the binding energy of the 4-particle complex with respect to the exciton plus a frozen FS increases steadily from $0.47 R_X$ to $0.60 R_X~ (0.58 R_X)$  as $k_F a_X$ increases from 0 to 0.7 for  2D QW,  and from $0.19 R_X$ to $0.22 R_X~ (0.20 R_X)$  as $k_F a_X$ increases from 0 to 0.5 for  quasi-2D QW. This feature can be attributed to the increase in correlation energy as the FS hole gains more available states within the FS when $k_F$ increases. This energy gain then overcomes the reduction of the trion binding energy caused by Pauli blocking.
When many FS electron-hole pairs are included,  the nature of this state will totally change. However, this regime as expected for high doping  is beyond the scope of this paper.\

The cross-over from trion-hole complex to exciton-polaron can be seen from photo-absorption experiments. The final state $|f\ran$ in Eq.~(\ref{AvQp}) then is the $|\Psi^{(eeh)}_i\ran$ state of Eq.~(\ref{dTr}). The photo-absorption spectrum associated with the trion-hole complex or the exciton-polaron follows from
\bea
 A(\omega) &\propto& \sum_i \big|\lan \Psi^{(eeh)}_i |\sum_\vk  a^\dag_{\vk,-\frac{1}{2}}b^\dag_{-\vk+\vQ_p,\frac{3}{2}} |F_{N_+,N_-}\ran \big|^2  \nn\\
&& \times  |\hat{\bf e}_p \cdot {\bf P}_{cv}|^2  \delta(\hbar \omega- E_g -{\cal E}^{(eeh)}_{i}) \label{app:absorp}\\
&\approx& \sum_i \frac { |\hat{\bf e}_p \cdot{\bf P}_{cv}|^2|\sum_{n,j}z_{X,j}^{(i)}x^{(j)}_{n,0} C_{n,0}\tilde f_{n,0}(0) |^2}{(\hbar \omega- E_g -
{\cal E}^{(eeh)}_i)^2+\gamma^2 }  \,,\nn
\eea
where $\tilde f_{n,0}(r)$ is the radial part of the $m=0$ exciton basis state defined in Eq.~(\ref{tildefnm1}) with $\tilde f_{n,0}(0)=1$ for polarized FS and $\tilde f_{n,0}(0)=1-\int^{k_F}_0 kdk I_0(\alpha_n,k)=\alpha_n/\sqrt{\alpha_n^2+k_F^2}$ for unpolarized FS.  $E_g$ is  the semiconductor band gap, $\hat{\bf e}_p$ is the photon polarization vector, and $C_{n,0}$ is the normalization factor of the exciton basis state $|n,0\ran$ as defined in Eq.~(\ref{Omn'n}). ${\bf P}_{cv}$ is the matrix element of the  momentum operator between the valence and conduction band Bloch states, which is essentially $\vk$-independent for the range of $\vk$ considered in the present calculation.  $z_{X,j}^{(i)}$ is the  prefactor of the $j$ exciton in the $|\Psi^{(eeh)}_i\ran$ state given in Eq.~(\ref{dTr}), and $x^{(j)}_{n,0}$ is the prefactor of the $|n,0\ran$ basis state in the exciton state $j$ given in Eq.~(\ref{X0}). Results are presented with the $\delta(x)$ peak replaced by $1/(x^2+\gamma^2)$ where $\gamma$ is a phenomenological absorption-line broadening taken to be equal to $0.015 R_X$.\ 

\begin{figure}[t]
\centering
\subfigure[]{\label{fig:9a} \includegraphics[trim=0.5cm 0cm 0.cm 0.5cm,clip,width=3in]  {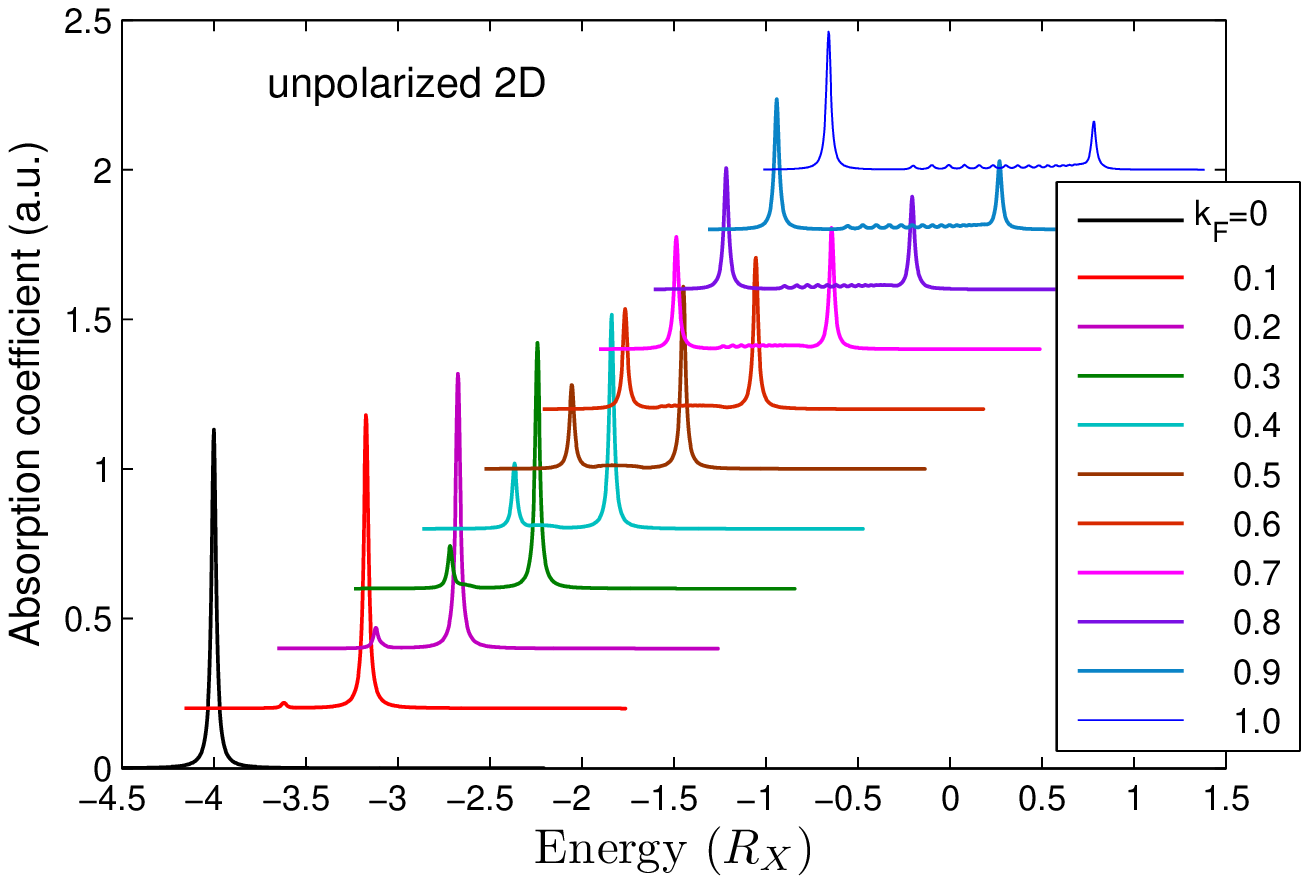}}
\subfigure[]{\label{fig:9b}  \includegraphics[trim=0.5cm 0cm 0.cm 0.5cm,clip,width=3in]  {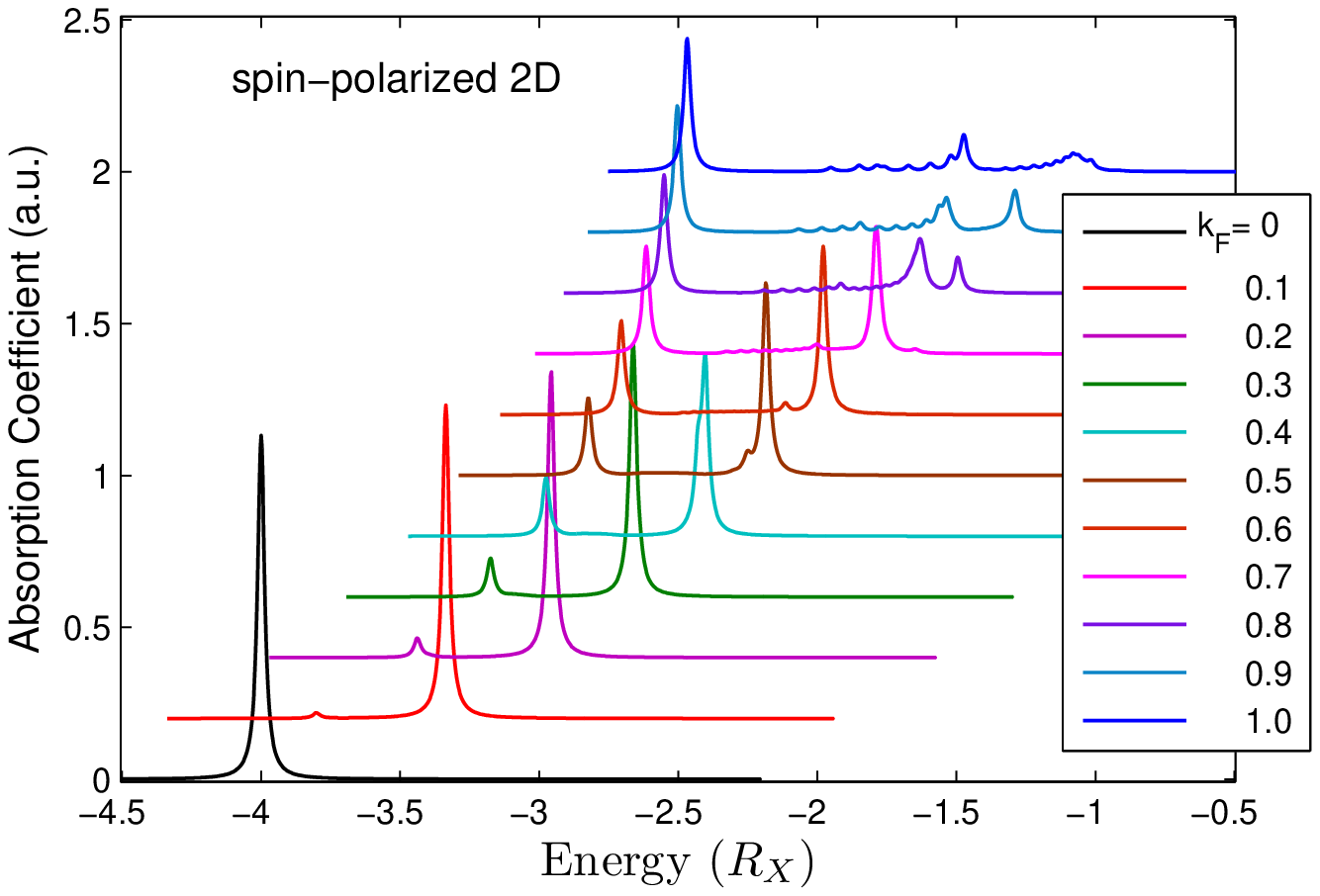}}
\caption{\small Absorption spectra of a photocreated electron-hole pair in the presence of (a) an unpolarized FS and (b) a spin-polarized FS for various  $k_F$ values (in units of $a_X^{-1}$), taking into account all possible excitations of single FS electron-hole pair for  2D QW. For clarity, the base lines of spectra with increasing $k_F$ values are vertically shifted up by $0.2$ from the pervious curve.}
\label{fig:9}
\end{figure}

\begin{figure}[t]
\centering
\subfigure[]{\label{fig:10a} \includegraphics[trim=0.5cm 0cm 0.cm 0.5cm,clip,width=3in]  {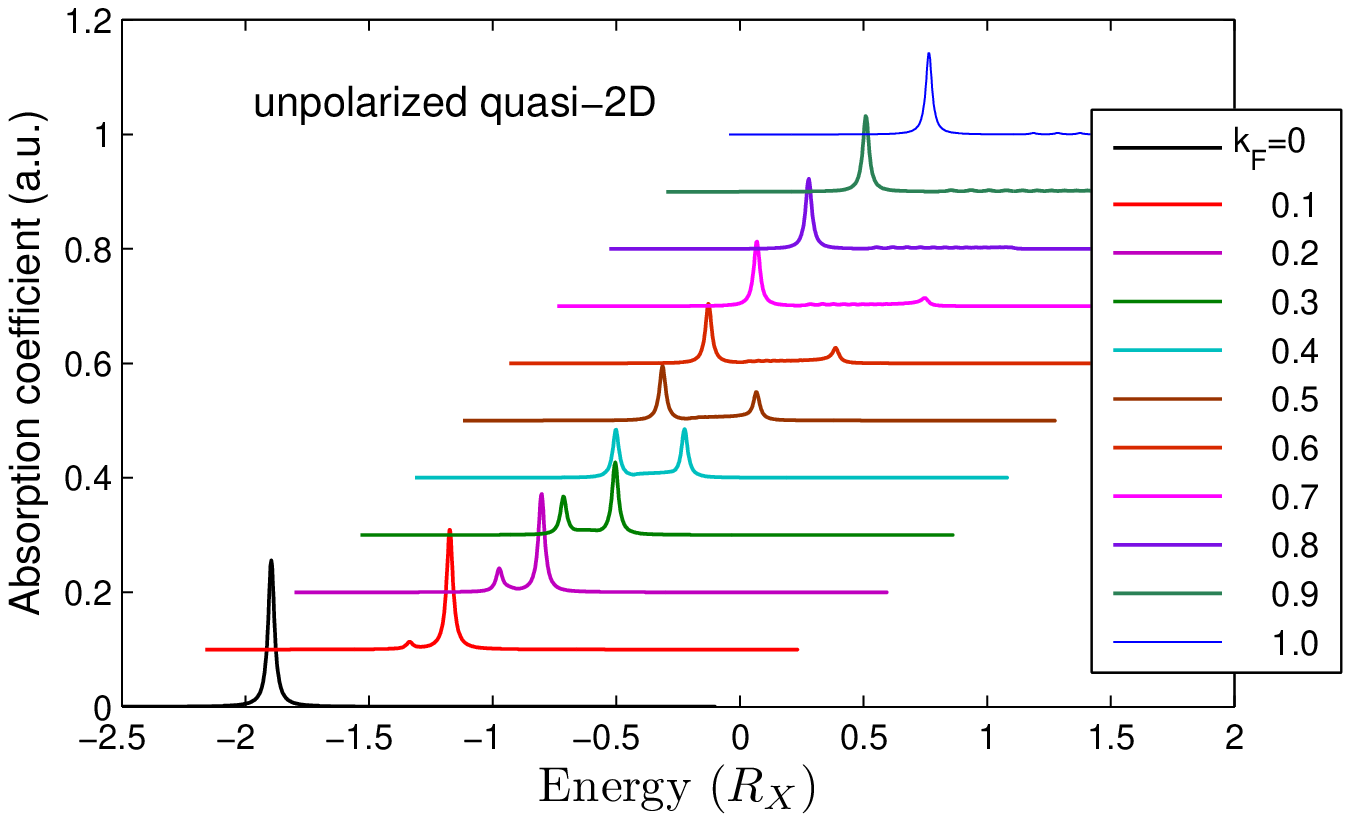}}
\subfigure[]{\label{fig:10b} \includegraphics[trim=0.5cm 0cm 0.cm 0.5cm,clip,width=3in]  {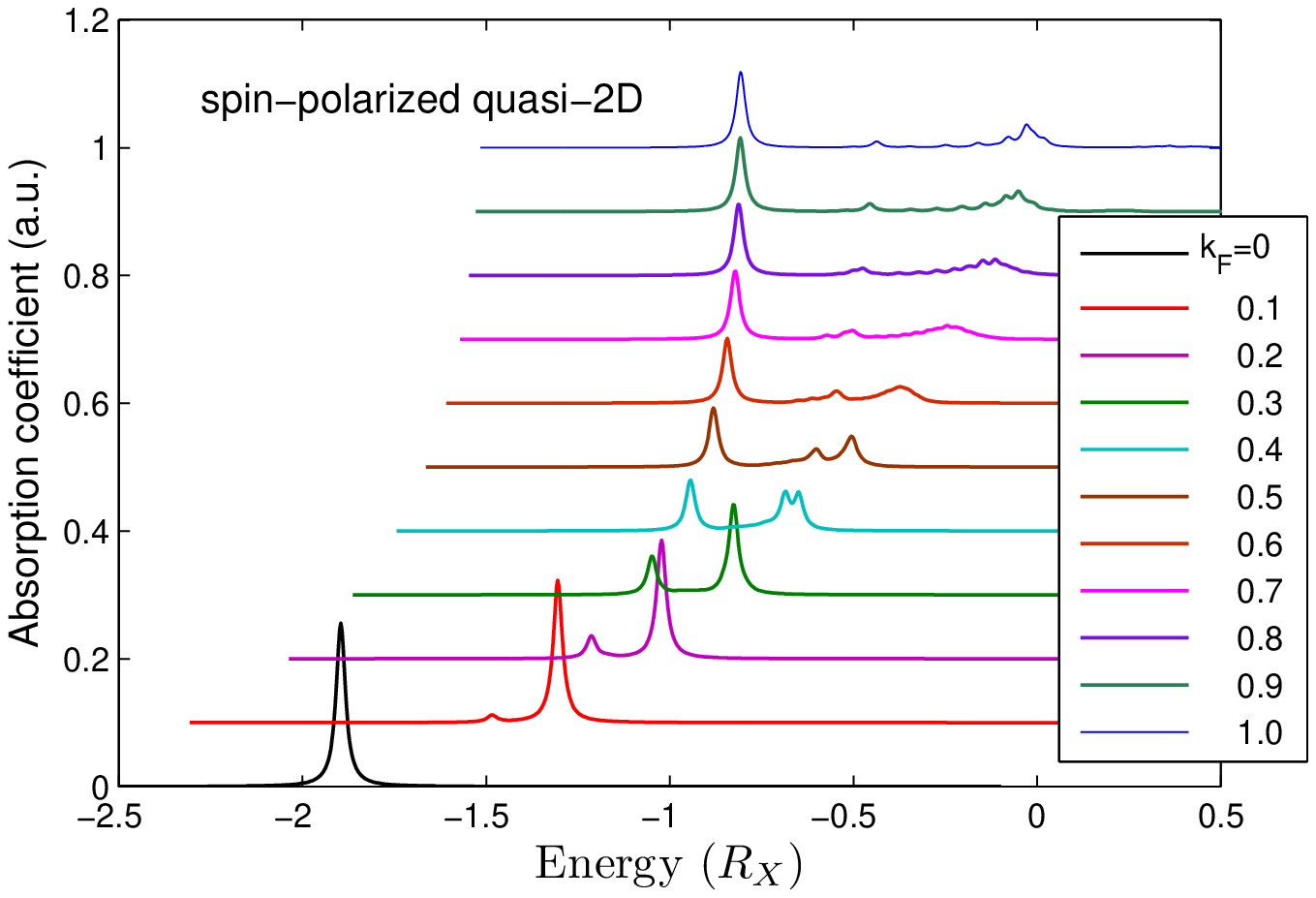}}
\caption{\small Same as Fig.~\ref{fig:9}, but for quasi-2D QW. For clarity, the base lines of spectra with increasing $k_F$ values are vertically shifted up by $0.1$ from the pervious curve.}
\label{fig:10}
\end{figure}

The calculated absorption spectra of doped semiconductors, as a function of $k_F$ (in units of $a_X^{-1}$), are shown in Figs.~\ref{fig:9} and \ref{fig:10} for  2D and quasi-2D, respectively. For clarity, the spectra for  different Fermi momenta $k_F$ are vertically shifted by $0.2$.   For $k_F=0$, only the exciton ground state peak appears, as the first excited state occurs at $-(4/9)R_X$ ($-0.34 R_X$) for 2D (quasi-2D), which is beyond the energy range of interest. As $k_F$ increases, another peak appears at the ground-state energy of the 4-particle complex, whose strength increases as $k_F$ increases, while the strength of the higher-energy peak gradually reduces. Note that the oscillator strength of each absorption peak is directly related to the amount of admixture of the exciton component in that state, since only the exciton is coupled directly to  photon.

At low doping, the 4-particle ground state corresponds to a trion weakly dressed by a conduction FS hole, and it turns into an exciton dressed by a  FS electron-hole pair as the doping density passes the crossing point, as evidenced by the increasing of oscillator strength. The cross-over occurs for spin-polarized (unpolarized) Fermi sea at $k_Fa_X\simeq 0.81 ~(0.72)$  in 2D QW and at $k_Fa_X\simeq 0.45 ~(0.39)$ in quasi-2D QW. At the crossing point, the two peaks have nearly the same strength (see  Figs.~\ref{fig:9} (Fig.~\ref{fig:10})).  Before the crossing point, the second peak remains exciton-like,  sitting at approximately $0.5R_X$ ($0.2R_X$) above the 4-particle ground state for  2D (quasi-2D) QW.

Beyond this crossing point, the exciton state enters the continuum band of the trion-hole excited states as indicated by the shaded area in Fig.~\ref{fig:8}. The coupling of the exciton with the group of trion-hole states leads to an anti-crossing behavior in much the same manner as in an exciton-polariton, with the ``lower branch" being exciton like and the upper branch being trion-hole like. The tracing of the energy of the upper branch of the ``coupled mode" is illustrated by solid  spheres for unpolarized FS in Fig.~\ref{fig:8}. It shows that the energy separation between the two main peaks grows wider as $k_F$ continues to increase beyond the crossing point.\

For spin-polarized FS, there is an overlap of  the trion-hole band associated with the trion ground state (lowest shaded area of  Fig.~\ref{fig:6}) with the trion-hole band associated with the trion first excited state (next shaded area in  Fig.~\ref{fig:6}) when the energy difference between the trion first-excited state and ground state becomes smaller than the band width ($E_F$) at high enough $k_F$ values. Such overlap produces some extra peaks in the absorption spectrum, as seen in Figs.~\ref{fig:9}(b) and \ref{fig:10}(b) for spin-polarized FS, since both groups of states can pick up some oscillator strength via coupling to the exciton state when their energies are close. The interplay of the two trion-hole bands coupling simultaneously to the exciton state after the cross-over point produces a rich anti-crossing pattern.\

To compare our theoretical predictions with experimental observations for II-VI semiconductors, we focus on the quasi-2D QW with unpolarized FS (see Fig.~\ref{fig:10}(a)). We compare our results with the reflectivity measurements of ZnSe/Zn$_{0.89}$Mg$_{0.11}$S$_{0.18}$Se$_{0.82}$ QW of 8nm well width reported in Ref.~\onlinecite{Suris}. Here, we take $R_X=22$meV and $a_X=3.3$nm, which correspond to the exciton binding energy and the effective Bohr radius in 3D ZnSe. For the quasi-2D Coulomb potential used here, we obtain an exciton binding energy of $1.896 R_X=42$meV and a trion binding energy of $0.188 R_X=4.1$meV when $E_F=0$, in good agreement with the experimental values reported in Ref.~\onlinecite{Suris}. As $E_F$ increases to 1.7meV, which corresponds to a value of $k_Fa_X\approx 0.3$, the observed energy splitting $\Delta_{XT}$ between the ground-state peak (trion like) and the excited-state peak (exciton like)  increases to  6meV (see Fig.~\ref{fig:4}(a) of Ref.~\onlinecite{Suris}), while our predicted value for this energy difference increases from $0.188 R_X=4.1$meV  at $E_F=0$ to $0.209 R_X=4.6$meV at $E_F=1.7$meV.  So, our model calculation predicts the increase of the  $\Delta_{XT}$ splitting with doping but underestimates the amount of increase observed experimentally. However, as $k_Fa_X$ increases to 0.4 (near cross-over point), our predicted value for $\Delta_{XT}$ quickly goes up to $0.278 R_X=6.1$meV.  Our model considers only basis states with angular functions up to $m=2$. By adding more basis functions, we expect the  $\Delta_{XT}$  splitting to be further increased, in closer agreement with experiment. Furthermore, the electron-to-hole mass ratio considered here is zero. A larger  electron-to-hole mass ratio would also increase the electron-hole correlation.\

 Overall, our results agree qualitatively with  experimental observations for many II-VI semiconductor QWs and  2D materials.  The energy splitting of the first two absorption peaks (labeled as $\Delta_{XT}$) increases as the doping density increases. It should be stressed that the so-called  $\Delta_{XT}$ splitting represents the energy difference between the exciton peak and trion state only at $k_F=0$. For finite $k_F$, such a splitting represents the energy difference in the two coupled modes of the exciton-polaron and the trion-hole complex. Furthermore, the cross-over behavior predicted here implies that the oscillator strength of the ground-state peak would increase as it picks more admixture of the exciton state, while the excited-state  peak would decrease as the doping density increases. This behavior has been observed experimentally and also predicted theoretically by many-body Green's function approach as reported in Ref.~\onlinecite{Suris}.\

\mbox{}

\section{Conclusion}

In this work, we study the absorption of a circularly-polarized ($\sigma_+$) photon in the presence of an unpolarized or fully polarized Fermi sea made of electrons having a spin different from  the photocreated electron.
We only consider single-pair excitations from the FS as induced by Coulomb interaction with the photoexcited exciton, that is, a 4-particle system. We show that  at low doping, its ground state essentially corresponds to a trion  weakly bound to a FS hole. When the doping increases,  the ground state turns into an exciton-polaron because of the increasing effect of Pauli blocking on the trion-hole complex. The cross-over from trion-hole to exciton-polaron occurs  for unpolarized (spin-polarized) FS at $k_Fa_X =$ 0.7 (0.8) for 2D QW, and at $k_Fa_X =$ 0.4 (0.45) for  quasi-2D QW. For a photon with circularly-polarized  polarization the photo-absorption spectra show two prominent low-energy peaks that correspond to the coupled states of the trion-hole complex and the exciton-polaron.
Their line shapes will be further modified by including more electron-hole pairs in the FS, which will be important in high doping regime.
By contrast, if the absorbed photon has a $\sigma_-$  polarization, the photocreated electron has  spin $(1/2)$. As trion cannot be formed with a polarized FS made of spin-$(1/2)$ electrons, we then have a pure exciton absorption spectrum, the FS possibly leading to  Fermi edge singularity. As a result, when using an unpolarized light source, we must see, in addition to the above two peaks associated with  exciton--trion-hole coupled states, a higher-energy peak that corresponds to the pure exciton state.

\acknowledgments
This work was supported under Contract No. MOST 107-2112-M-001-032. S.Y. S. would like to thank INSP, CNRS for invitations.

\begin{widetext}

\renewcommand{\thesection}{\mbox{Appendix~\Roman{section}}} 
\setcounter{section}{0}

\renewcommand{\theequation}{\mbox{A.\arabic{equation}}} 
\setcounter{equation}{0} %
\section{Relevant matrix elements for exciton states\label{app:sec1}}

\textbf{(1) In the absence of  Fermi sea}, the overlap between the conduction electron-valence hole pair states $|n,m\ran$ defined in Eq.~(\ref{X1}) reads as
\be \lan n',m  |n,m\ran=O^{(m)}_{n',n}= C_{n',m}C_{n,m} 2\pi \frac{(3-2\delta_{m,0})!}{(\alpha_n+\alpha_{n'})^{4-2\delta_{m,0}}}\,
\ee
with $C_{n',m}=(2\alpha_n)^{2-\delta_{m,0}}/\sqrt{2\pi(3-2\delta_{m,0})}$. The kinetic energy part of the Hamiltonian is given by
\bea
\lan n',m |H_0 |n,m\ran &=& K^{(m)}_{n',n} = 2\pi C_{n',m} C_{n,m}\int_0^\infty dr ~r^{2-\delta_{m,0}} e^{-\alpha_{n'} r}\frac{1}{2m^*_e}\left(\frac{m^2}{r^2}-\frac{\partial}{r\partial r}\left(r\frac{\partial}{\partial r}\right)\right)r^{1-\delta_{m,0}}e^{-\alpha_n r}\nn\\
&=&2\pi C_{n',m} C_{n,m}\frac{1}{2m^*_e}\left(\frac{m^2-(1-\delta_{m,0})^2}{(\alpha_n+\alpha_{n'})^{2-2\delta_{m,0}}}
+\frac{\alpha_n\alpha_{n'}(3-2\delta_{m,0})!}{(\alpha_n+\alpha_{n'})^{4-2\delta_{m,0}}}\right)
\, , \label{KE}
\eea
while the electron-hole Coulomb  part is given by
\bea
\lan n',m |V_{e h} |n,m\ran
&=&-  2\pi C_{n',m}  C_{n,m} \int d r r^{3-2\delta_{m,0}} e^{-(\alpha_n+\alpha_{n'})r} v(\vr)\nn \\
&=& - C_{n',m}  C_{n,m}\frac {2\pi e^2}{\epsilon_{sc}}\frac{2-\delta_{m,0}}{(\alpha_n+\alpha_{n'})^{3-2\delta_{m,0}}}\, .
\eea

\textbf{(2) In the presence of a Fermi sea having a Fermi wave vector $k_F$}, wave functions in momentum space, as the basis functions defined in Eq.~(\ref{Fm}), are more convenient to take into account Pauli blocking.  Let us write them as
$F_{n,m}(k)=2\pi (-i)^m I^{(m)}_{1-\delta_{m,0}}(\alpha_n, k)$, with
\be
I^{(m)}_{L}(\alpha,k)= \left(-\frac{\partial}{\partial \alpha}\right)^{L+1} \left[\frac{(f(\alpha)-\alpha)^{|m|}}{k^{|m|}f(\alpha)}\right],
\ee
where $f(\alpha)=\sqrt{\alpha^2+k^2}$.\

For $m=0$, we have
$I^{(0)}_0(\alpha,k)=\alpha/ f^3$ and $I^{(0)}_2(\alpha,k)=\alpha(15\alpha^2-9f^2)/f^7$, while  for $m>0$,
\be
I^{(m)}_{1}(\alpha,k)=(f-\alpha)^m[(m+3f')(mf+\alpha)-m\alpha-f]/(k^mf^4)
\ee
and
\bea
I^{(m)}_{2}(\alpha,k) &= & (f-\alpha)^{(m-1)}[4(f-\alpha)f'-m(\alpha-f)][(m+3f')(mf+\alpha)-m\alpha-f]/(k^mf^5) \nn \\
& & -(f-\alpha)^m[(m+3f')(mf'+1)+3(1/f-\alpha f'/f^2)(mf+\alpha)-m-f']/(k^mf^4)\, ,
\eea
where $f'=\alpha/f$.

 The overlap then reads as
\be
O^{(m)}_{n',n}= C_{n',m} C_{n,m} \frac{1}{2\pi} \int_{k_F}^{\infty}kdk ~ F^*_{n',m}(k)F_{n,m}(k)\,,\label{Omn'n}
\ee
with
\be C_{n,m}=1/\sqrt{ \frac{1}{2\pi} \int_{k_F}^{\infty}kdk |F_{n,m}(k)|^2}\,. \ee
We have
\be
O^{(0)}_{n',n}= C_{n',0}  C_{n,0} \pi\alpha_{n'}\alpha_n J^{(0)}_{n',n}\,,
\ee
\be
O^{(1)}_{n',n}= C_{n'1}  C_{n,1} 9\pi\alpha_{n'}\alpha_n M^{(1)}_{n',n}\,,
\ee
\be
O^{(2)}_{n',n}= C_{n'2}  C_{n,2} 9\pi\alpha_{n'}\alpha_n M^{(2)}_{n',n}\,,
\ee
while the kinetic energy part of $\lan n',m |H |n,m\ran$ reads as
\be
K^{(m)}_{n',n}= C_{n',m}  C_{n,m} \frac{1}{2\pi} \int_{k_F}^{\infty}k^3dk~F^*_{n',m}(k)F_{n,m}(k)\,,
\ee
with
\be
K^{(0)}_{n',n}=C_{n',0}  C_{n,0} \pi\alpha_{n'}\alpha_n J^{(1)}_{n',n},
\ee
\be
K^{(1)}_{n',n}= C_{n',1}  C_{n,1} 9\pi\alpha_{n'}\alpha_n M^{(2)}_{n',n}\,,
\ee
\be
K^{(2)}_{n',n}= C_{n',2}  C_{n,2} 9\pi\alpha_{n'}\alpha_n M^{(3)}_{n',n}\,,
\ee
where
\be
J^{(m)}_{n',n}=\int_{k_F^2}^{\infty} du \frac {u^m} {\sqrt{(a+bu+u^2)^3}}\,,
\ee
\be
M^{(m)}_{n',n}=\int_{k_F^2}^{\infty} du \frac {u^m} {\sqrt{(a+bu+u^2)^5}}\,,
\ee
with $a=(\alpha_{n'}\alpha_n)^2$ and $b={\alpha_{n'}}^2+\alpha_n^2$. Explicit expressions of  $J^{(m)}_{n',n}$ and $M^{(m)}_{n',n}$ for the  first few $m$'s are
\bea
J^{(0)}_{n',n}&=&-\frac 4 {\Delta} + \frac {2(2k_F^2+b)}{\Delta \sqrt{a+bk_F^2+k_F^4}} \longrightarrow \frac 2 {\alpha_{n'}\alpha_n(\alpha_{n'}+\alpha_n)^2} \mbox{ as } k_F \rightarrow  0, \\
J^{(0)}_{n,n}&=&\frac 1 {2(\alpha_n^2+k_F^2)^2} ,\\
J^{(1)}_{n',n}&=&\frac {2b} {\Delta} + \frac {2(2a+bk_F^2)}{\Delta \sqrt{a+bk_F^2+k_F^4}} \longrightarrow \frac 2 {(\alpha_{n'}+\alpha_n)^2} \mbox{ as } k_F \rightarrow  0, \\
J^{(1)}_{n,n}&=&\frac {(\alpha_n^2/2+k_F^2)} {(\alpha_n^2+k_F^2)^2} .
\eea
\bea
M^{(0)}_{n',n}&=&  \frac {2(2k_F^2+b)} {3\Delta \sqrt{(a+bk_F^2+k_F^4)^3}}-\frac {16}{3\Delta} (\frac {2k_F^2+b}{\Delta \sqrt{a+bk_F^2+k_F^4}}-\frac 2 {\Delta} )
\longrightarrow \frac 2 {\alpha_{n'}\alpha_n(\alpha_{n'}+\alpha_n)^2} \mbox{ as } k_F \rightarrow  0, \\
M^{(0)}_{n,n}&=&\frac 1 {4(\alpha_n^2+k_F^2)^4} ,\\
M^{(1)}_{n',n}&=& \frac 1 {3\sqrt{(a+bk_F^2+k_F^4)^3}} - \frac b 2 M^{(0)}_{n',n}
\longrightarrow \frac 4 {3\alpha_{n'}\alpha_n(\alpha_{n'}+\alpha_n)^4} \mbox{ as } k_F \rightarrow  0,\\
M^{(2)}_{n',n}&=& \frac {k_F^2}{2\sqrt{(a+bk_F^2+k_F^4)^3}}  - \frac b 4 M^{(1)}_{n',n} + \frac a 2 M^{(0)}_{n',n}
\longrightarrow \frac 4 {3\alpha_{n'}\alpha_n(\alpha_{n'}+\alpha_n)^4} \mbox{ as } k_F \rightarrow  0,\\
M^{(3)}_{n',n}&=& \frac {k_F^4}{\sqrt{(a+bk_F^2+k_F^4)^3}}+2a M^{(1)}_{n',n} + \frac b 2 M^{(2)}_{n',n} \longrightarrow\frac {2(b+4\alpha_{n'}\alpha_n)} {3\alpha_{n'}\alpha_n(\alpha_{n'}+\alpha_n)^4} \mbox{ as } k_F \rightarrow  0.
\eea
with $\Delta=b^2-4a=(\alpha_{n'}^2-\alpha_n^2)^2$.

The matrix elements for the electron-hole Coulomb potential read
\bea
\lan n',m |\tilde V_{eh} |n,m\ran
&=& - C_{n',m} C_{n,m}  \int d{\bf r}  \tilde f_{n',m}(r) \tilde f_{n,m}(r) \tilde V^{sc}(r),  \, \label{PE1}
\eea
where
\be
\tilde f_{n,m}(r)=f_{n,m}(r)-\int_0^{k_F} k dk J_m(kr)I_{1-\delta_{m,0}}^{(m)}(\alpha_n,k)\label{tildefnm1}
\ee
is the radial part of the FS-blocked basis function in real space, and
\be \tilde V^{sc}(r)= \frac{ e^2}{\epsilon_{sc}}\left[ \frac {Z(r/r_0)} {r}-\frac{2\pi}{ A}\sum_{\vq}\frac {s_q}{q(1+qr_0)(q(1+qr_0)+s_q)} e^{i\vq\cdot \vr} \right],  \, \label{Vsc}
\ee
where the first term on the right-hand side (RHS) is the bare Coulomb potential and the second term is the screening part which can be evaluated accurately, since the integrand decays quickly with $q$.
\mbox{}

\renewcommand{\theequation}{\mbox{B.\arabic{equation}}} 
\setcounter{equation}{0} %
\section{Matrix elements  for the electron-electron Coulomb potential\label{app:sec2}}

Since we have excluded the Coulomb interactions with the remaining $N$ electrons, the semiconductor Hamiltonian $H$ is reduced to $H_T+\mathcal{E}_{N}$,
where $H_T$ denotes the Hamiltonian of the three particles $(h,e_1,e_2)$. For infinite valence hole mass, we have
\be
H_T= H_1+H_2+V_{e_1e_2} \,,
\ee
with $H_j=\va^{(e_j)}_\vk-V_{e_jh}$. As $H_1$ does not involve the $e_2$ electron, its matrix element $\lan n'_1,n_2',m'|H_1|n_1,n_2,m\ran$ reads as $\delta_{m',m}\delta_{n_2',n_2}\lan n'_1,m| H_1|n_1,m\ran$; similarly for $\lan n'_1,n_2',m'|H_2|n_1,n_2,m\ran$. These two matrix elements are orthogonal with respect to angular momentum $m$.
On the other hand, the matrix element for $V_{e_1e_2}$ is given by
\bea
\lefteqn{ \lan n'_1,n_2',m' |V_{e_1e_2} |n_1,n_2,m \ran =  C_{n_1', m'}  C_{n'_2, m'}  C_{n_1, m} C_{n_2, m}} \nn\hspace{2cm} \\
&&\times  \int d\vr_1 \phi^*_{n'_1,m'}(\vr_1)\phi_{n_1,m}(\vr_1) \int d\vr_2 \phi^*_{n'_2,-m'}(\vr_2)\phi_{n_2,-m}(\vr_2) \tilde  V^{sc}(|\vr_1-\vr_2|)\,
\eea
where $ \tilde  V^{sc}(r)$ is given in Eq.~(\ref{Vsc}).
We end up with
\bea
\lefteqn{ \lan n'_1,n_2',m' |V_{e_1e_2} |n_1,n_2,m \ran =  C_{n_1', m'}  C_{n'_2, m'}  C_{n_1, m} C_{n_2, m} }\hspace{2cm} \nn \\
&&\times 4\pi^2 \int r_1dr_1 \tilde f_{n'_1,m'}(r_1)\tilde f_{n_1,m}(r_1) \int r_2 dr_2 \tilde f_{n'_2,m'}(r_2)\tilde f_{n_2,m}(r_2) \bar  V^{sc}_{|m-m'|}(r_1,r_2)\, ,
\eea
where
\be
\bar V^{sc}_{|m-m'|}(r_1,r_2)=\int_0^{\pi} \frac {d\phi}{\pi} cos(|m-m'|\phi) \tilde V^{sc}(\sqrt{r_1^2+r_2^2-2r_1r_2\cos\phi}).\label{app:B4}
\ee

It should be noted that the potential $ \tilde V^{sc}(r_{12})$ contains a $1/r_{12}$ singular term in the bare-potential part, which diverges as $r_1 \rightarrow r_2$. Thus, care must be exercised when performing the real-space integration. We rewrite the integral over $r_1$ ad $r_2$ in Eq.~(\ref{app:B4}) as
\bea
\lefteqn{ \int r_1dr_1 \tilde f_{n'_1,m'}(r_1)\tilde f_{n_1,m}(r_1) \int r_2 dr_2 \tilde f_{n'_2,m'}(r_2)\tilde f_{n_2,m}(r_2) \bar  V^{sc}_{|m-m'|}(r_1,r_2) }\hspace{2cm}  \nn \\
& =& \int r_> dr_>  \int_0^{r_>} r_< dr_< \Big[\tilde f_{n'_1,m'}(r_>)\tilde f_{n_1,m}(r_>) \tilde f_{n'_2,m'}(r_<)\tilde f_{n_2,m}(r_<)   \nn \\
&& +\tilde f_{n'_1,m'}(r_<)\tilde f_{n_1,m}(r_<) \tilde f_{n'_2,m'}(r_>)\tilde f_{n_2,m}(r_>)\Big] \bar  V^{sc}_{|m-m'|}(r_<,r_>) \nn \\
& =& \int r_>^3 dr_>  \int_0^{1} p dp \Big[\tilde f_{n'_1,m'}(r_>)\tilde f_{n_1,m}(r_>) \tilde f_{n'_2,m'}(p~r_>)\tilde f_{n_2,m}(p~r_>)  \nn \\
&& +\tilde f_{n'_1,m'}(p~r_>)\tilde f_{n_1,m}(p~r_>) \tilde f_{n'_2,m'}(r_>)\tilde f_{n_2,m}(r_>)\Big] \bar  V^{sc}_{|m-m'|}(r_>,p~r_>),
\eea
where $r_{>}=max(r_1,r_2)$ and $r_{<}=min(r_1,r_2)$ .
For the singular part of $ \tilde V^{sc}(r_{12})$, the angular integration given in Eq.~(\ref{app:B4}) reduces to elliptic functions of the ratio $p=r_</r_>$.

In the limit $k_F \rightarrow 0$,  we have
\bea
\lefteqn{\lim_{k_F\rightarrow 0} \lan n'_1,n'_2,m' |V_{e_1e_2} |n_1,n_2,m \ran= \frac{4\pi^2 e^2}{\epsilon_{sc}}  C_{n'_1,m'} C_{n'_2,m'} C_{n_1,m}C_{n_2,m}}\hspace{2cm}   \nn \\
&&\times \int  dq   I^{(m-m')}_{2-\delta_{m,0}-\delta_{m',0}}(\alpha_{n'_1}+\alpha_{n_1},q)  I^{(m-m')}_{2-\delta_{m,0}-\delta_{m',0}}(\alpha_{n'_2}+\alpha_{n_2},q).
\eea

\renewcommand{\theequation}{\mbox{C.\arabic{equation}}} 
\setcounter{equation}{0} %
\section{Resolution of  Eq.~(\ref{Schrod:dTr})\label{app:sec3}}

\mbox{}

$\bullet$  {\bf Polarized Fermi sea}

\mbox{}

We first project $(H-E^{(eeh)}_i)|\Psi^{(eeh)}_i\ran=0$  into the exciton state $|\Psi^{(e)}_{j'}\ran$ with $m=0$. This leads to
\bea
\lefteqn{ ( E^{(e)}_{j'}-E^{(eeh)}_i) z^{(i)}_{X,j'}+ \sum_{j} \sum_{\vq,\ell}  \lan \Psi^{(e)}_{j'} |H a_{\vq,\frac 1 2}|\Psi^{(ee)}_j\ran G_{\ell,0}(q)  z^{(i)}_{T,j,\ell} }\hspace{1cm} \nn \\
&+& \sum_{n_1,n_2,\ell,m_1,m_3} \lan \Psi^{(e)}_{j'} |H|n_1,n_2,\ell,m_1,m_3,1/2\ran z^{(i)}_{n_1,n_2,\ell,m_1,m_3,\frac{1}{2}} =  0\, ,\label{app:HamPol}
\eea
where $ E^{(e)}_{j'}$ are exciton  eigen-energies.
The second term describes the coupling between the exciton and  the trion-hole states (including bound and unbound states) that is induced by  Coulomb interaction. Here, $\vq$ and $\vq'$ are restricted in the spin-$(1/2)$ Fermi sea, $|F_{N_+}\ran$. We have
\bea
\lefteqn{  \lan \Psi^{(e)}_{j'}  | H    a_{\vq,\frac 1 2}|\Psi^{(ee)}_j\ran = \sum_{n_1,n_2,n'_1,m} x^{(j')}_{n'_1,0} t^{(j)}_{n_1,n_2,m} \sum_{\vk_2} \phi_{n_2,-m}(\vk_2) } \hspace{1cm}\nn \\
 &&\times   \Big[ \sum_{\vk'_1,\vk_1} \phi^*_{n'_1,0}(\vk'_1) \phi_{n_1,m}(\vk_1) v_{ee}(\vk_1,\vk_2;\vk'_1,\vq)-\delta_{m,0}O^{(0)}_{n_1,n'_1}  V(\vk_2-\vq)\Big]\, , \label{newf}
\eea
where the term $ v_{ee}(\vk_1,\vk_2;\vk'_1,\vq)$ comes from the Coulomb interaction between the electron $e_1$ with spin $(-1/2)$ in the exciton and the FS electron $e_2$ with spin  $(1/2)$; the term  $V(\vk_2-\vq)$ comes from the scattering of the FS electron  $e_2$  by the Coulomb interaction  with the valence hole.

The matrix elements  are more efficiently evaluated in real-space integrals. We have
\bea
\lefteqn{ \sum_{\vq,\ell,0}  \lan \Psi^{(e)}_{j'} |H a_{\vq,\frac 1 2}|\Psi^{(ee)}_j\ran G_{\ell,0}(q)= \sum_{n_1,n_2,n'_1,m} x^{(j')}_{n'_1,0} t^{(j)}_{n_1,n_2,m}  \Big[C_{n_1', 0}   C_{n_1,m} C_{n_2,m}  }\hspace{1cm} \nn \\
&&\times \int d\vr_1 \tilde f_{n'_1,0}(r_1)\tilde f_{n_1,m}(r_1) \int d\vr_2 \tilde f_{n_2,m}(r_2)Q_{\ell,0}(r_2)  \bar V^{sc}_{|m|}(r_1,r_2)+\delta_{m,0}O_{n'_1,n_1}\lan n_2,0|\tilde V_{eh}|\ell,0\ran \Big] \, ,
\eea
where
\be
\lan n'_2,m |\tilde V_{eh} |\ell,m\ran = - C_{n'_2,m}   \int d{\bf r}  \tilde f_{n'_2,m}(r) Q_{\ell,m}(r) \tilde V^{sc}(r),  \, \label{PE2}
\ee
and
\be
Q_{\ell,m}(r)=\frac{i^m}{2\pi} \int_0^{k_F} q dq G_{\ell,0}(q) J_{m}(qr).
\ee
For the third term in Eq.~(\ref{app:HamPol}), we have
\bea
\lefteqn{\lan \Psi^{(e)}_{j'} |H|n_1,n_2,\ell,m_1,m_3,1/2\ran \nn =\sum_{n'_1} x^{(j')}_{n'_1,0} C_{n_1', 0}   C_{n_1, m_1} C_{n_2, m_2}  }\hspace{1cm}\nn  \\
&& \times 4\pi^2\int r_1dr_1 \tilde f_{n'_1,0}(r_1)\tilde f_{n_1,m_1}(r_1) \int r_2dr_2 \tilde f_{n_2,m_2}(r_2) Q_{\ell,m_3}(r_2) \bar V^{sc}_{|m_1|}(r_1,r_2) \nn  \\
&&+ \delta_{m_1,0}\sum_{n'_1} x^{(j')}_{n'_1,0}O^{(0)}_{n'_1,n_1}\lan n_2,m_3|\tilde V_{eh}|\ell,m_3\ran  \, ,
\eea
where $m_2=-(m_1+m_3)$.
We next  project  the Schr\"{o}dinger equation for $|\Psi^{(eeh)}_i\ran$ into  the trion-FS hole basis state $ G_{\ell',0}(q')  a_{\vq'_1,\frac 1 2}|\Psi^{(ee)}_{j'}\ran$. This gives
\bea
 \sum_{\vq'} G_{\ell',0}(q')  \lan \Psi^{(ee)}_{j}|a^\dag_{ \vq',\frac{1}{2}} H  |\Psi^{(e)}_{j}\ran z^{(i)}_{X,j}+ \sum_{j,\ell} \sum_{\vq',\vq}G_{\ell',0}(q') \lan \Psi^{(ee)}_{j'}|a^\dag_{ \vq',\frac 1 2} (H-E^{(eeh)}_i) a_{\vq,\frac 1 2}|\Psi^{(ee)}_{j}\ran G_{\ell,0}(q) z^{(i)}_{T,j,\ell} \nn \\
 + \sum_{m_3\ne 0} \sum_{n_1,n_2,\ell,m_1} \sum_{\vq'} G_{\ell',0}(q')  \lan \Psi^{(ee)}_{j}|a^\dag_{ \vq',\frac 1 2 } H  |n_1,n_2,\ell,m_1,m_3,\frac{1}{2}\ran z^{(i)}_{n_1,n_2,\ell,m_1,m_3,\frac{1}{2}} =  0 \, .
\eea
where it is understood that the sums over $\vq$ and $\vq'$ are restricted inside the FS due to the cut-off function included in  $G_{\ell,0}(q)$.
The matrix elements in the first term  of the above equation are written explicitly as
\bea
\lefteqn{ \sum_{\vq'\vq}   G_{\ell',0}(q')\lan \Psi^{(ee)}_{j'}|a^\dag_{ \vq',\frac 1 2 } (H-E^{(eeh)}_i) a_{\vq,\frac 1 2}|\Psi^{(ee)}_{j}\ran  G_{\ell,0}(q)= }\nn\hspace{1cm} \\
&&  \Big[( E^{(ee)}_j-E^{(eeh)}_i)\delta_{\ell',\ell}-\tilde K_{\ell',\ell}+  \sum_{\vq,\vq'}  G_{\ell',0}(q')V(\vq'-\vq) G_{\ell,0}(q)\Big]\delta_{j',j}   \nn\\
&&- \sum_{n_1,n_2,m}t^{(j')}_{n'_1,n'_2,m} t^{(j)}_{n_1,n_2,m}\sum_{\vq',\vq}  G_{\ell',0}(q') \Big[O^{(m)}_{n_2',n_2}\sum_{\vk'_1,\vk_1} \phi^*_{n',m}(\vk'_1) \phi_{n,m}(\vk_1)v_{ee}(\vk'_1,\vq;\vk_1,\vq') \nn\\
&& + O^{(m)}_{n'_1,n_1}
 \sum_{\vk'_2,\vk_2}\phi^*_{n_2',-m}(\vk'_2) \phi_{n_2,-m}(\vk_2)\big(v_{ee}(\vk'_2,\vq;\vk_2,\vq')-v_{ee}(\vk'_2,\vq;\vq',\vk_2)\big)
\Big]G_{\ell,0}(q) \, , \label{newG}\eea
where  $E^{(ee)}_j$ is the $j$ trion-state energy.

On the RHS of the above equation, the first term describes the energy for the trion part. The  second term contains the kinetic-energy matrix elements for the FS hole defined by
\be \tilde K_{\ell',\ell}=\frac 1 {2\pi} \int_0^{k_F}q^3 dq  G_{\ell',0}(q)G_{\ell,0}(q),\ee
which can be carried out analytically and we have
$\tilde K_{\ell,\ell}= k_F^2(1/3-1/(2\ell\pi)^2)$ and $\tilde K_{\ell',\ell}= (k_F/\pi)^2[1/(\ell'-\ell)^2-1/(\ell+\ell')^2](-1)^{\ell+\ell'}$ for $\ell'  \ne \ell$. The third term is the Coulomb interaction between the valence hole and the FS hole, which is repulsive, whose matrix elements
are simply
\be
\lan \ell',0|V_{eh}|\ell,0 \ran =   \sum_{\vq,\vq'}G_{\ell',0}(q')V(\vq'-\vq)G_{\ell,0}(q)= \frac 1 {4\pi^2}  \int_0^{k_F} q'dq'  \int_0^{k_F}q dq G_{\ell',0}(q') \tilde V^{sc}_0(q',q)G_{\ell,0}(q), \label{Veh}
\ee
where
\be
\tilde V^{sc}_m(q',q)=\int_0^{\pi} \frac {d\phi}{\pi} cos(m\phi)\frac {2\pi e^2}{\epsilon_{sc}q(1+q r_0)\kappa_q}|_{q=\sqrt{q_1^2+q_2^2-2q_1q_2\cos\phi}}.
\ee
The fourth term contains the Coulomb interaction between the spin-$(1/2)$ electron of the trion and the FS hole, and finally the direct and exchange Coulomb interactions between the spin-$(1/2)$ electron of the trion and the FS hole.

The explicit matrix elements for direct Coulomb scattering between electron and FS-hole are
\bea
\lefteqn{ \sum_{\vk'_2,\vk_2,\vq',\vq}G_{\ell',0}(q') \phi^*_{n'_2,-m}(\vk'_2)v_{ee}(\vk'_2,\vq;\vk_2,\vq')   \phi_{n_2,-m}(\vk_2)G_{\ell,0}(q)}\hspace{1.5cm}  \nn \\
&=& C_{n'_2,m} C_{n_2,m} 4\pi^2
 \int r_2 dr_2 \int r_3dr_3 \tilde f_{n'_2,m}(r_2)\tilde f_{n_2,m}(r_2) \bar V^{sc}_{0}(r_2,r_3)  Q_{\ell',0}(r_3)Q_{\ell,0}(r_3) \, ,
\eea
and for the exchange Coulomb scattering
\bea
\lefteqn{ \sum_{\vk'_2,\vk_2,\vq',\vq}G_{\ell',0}(q') \phi^*_{n'_2,-m}(\vk'_2)  \phi_{n_2,-m}(\vk_2) v_{ee}(\vk'_2,\vq;\vq',\vk_2) G_{\ell,0}(q)} \hspace{1.5cm}\nn \\
&=& C_{n'_2,m} C_{n_2,m}4\pi^2
 \int r_2 dr_2\int r_3dr_3 \tilde f_{n'_2,m}(r_2)Q_{\ell',0}(r_3) \bar V^{sc}_{|m|}(r_2,r_3) Q_{\ell,0}(r_2)\tilde f_{n_2,m}(r_3)  \, .
\eea

Similarly, the matrix elements for the coupling between trion-hole states and the $m_3\ne 0$ basis states are given by
\bea \lefteqn{ \sum_{\vq'} G_{\ell',0}(q')  \lan \Psi^{(ee)}_{j}|a^\dag_{ \vq',\frac 1 2 } H  |n_1,n_2,\ell,m_1,m_3,\frac 1 2\ran } \nn \\
&=& -\sum_{n'_1,n'_2,\ell'} t^{(j)}_{n'_1,n'_2,m_2}O^{(m_2)}_{n'_2,n_2}  C_{n'_1,m_2} C_{n_1,m_1} \int d\vr_1 \int d\vr_3 \tilde f_{n'_1,m_2}(r_1)Q_{\ell',0}(r_3) \bar V^{sc}_{|m_3|}(r_1,r_3)\tilde f_{n_1,m_1}(r_1)Q_{\ell,m_3}(r_3) \nn \\
&&- \sum_{n'_1,n'_2,\ell'} t^{(j)}_{n'_1,n'_2,m_1} O^{(m_1)}_{n'_1,n_1}  C_{n'_2,m_1} C_{n_2,m_2}\Big[ \int d\vr_2  \int d\vr_3  \tilde f_{n'_2,m_1}(r_2)Q_{\ell',0}(r_3) \bar V^{sc}_{|m_3|}(r_2,r_3)\tilde f_{n_2,m_2}(r_2)Q_{\ell,m_3}(r_3)  \nn \\
&&- \int  d\vr_2\int d\vr_3 \tilde f_{n'_2,m_1}(r_2)Q_{\ell,m_3}(r_2) \bar V^{sc}_{|m_2|}(r_2,r_3) Q_{\ell',0}(r_3)\tilde f_{n_2,m_2}(r_3) \Big]|_{m_2=-m_1-m_3} \, .
\eea

Finally, we project  the Schr\"{o}dinger equation for $|\Psi^{(eeh)}_i\ran$ into  the $m_3\ne 0$ basis states. We obtain for the diagonal block associated with $m_3\ne 0$ basis states
\bea \lefteqn{ \lan n'_1,n'_2,\ell',m'_1,m'_3,\frac 1 2| H |n_1,n_2,\ell,m_1,m_3,\frac 1 2\ran =\delta_{m'_1,m_1} \delta_{m'_3,m_3} \Big\{ \big[K^{(m_1)}_{n'_1,n_1}+\lan n'_1,m_1|\tilde V_{eh}|n_1,m_1 \ran \big] O^{(m_2)}_{n'_2,n_2}\delta_{\ell',\ell} }  \nn \\
&&+ O^{(m_1)}_{n'_1,n_1}\big[ K^{(m_2)}_{n'_2,n_2}+\lan n'_2,m_2|\tilde V_{eh}|n_2,m_2 \ran \big]\delta_{\ell',\ell}
+O^{(m_1)}_{n'_1,n_1}O^{(m_2)}_{n'_2,n_2}\big[- \tilde K_{\ell',\ell}+\lan \ell',m_3|V_{eh}|\ell,m_3 \ran \big] \Big\}     \nn \\
&&+  \delta_{m'_3,m_3}\delta_{\ell',\ell} C_{n_1', m'_1}  C_{n'_2, m'_2}  C_{n_1, m_1} C_{n_2, m_2} \nn\hspace{1cm} \\
&& \times  \int d\vr_1 \tilde f_{n'_1,m'_1}(r_1)\tilde f_{n_1,m_1}(r_1) \int d\vr_2 \tilde f_{n'_2,m'_2}(r_2)\tilde f_{n_2,m_2}(r_2) \bar  V^{sc}_{|m_1-m'_1|}(r_1,r_2)\nn \\
&&-  \delta_{m'_2,m_2}O^{(m_2)}_{n'_2,n_2} C_{n'_1,m'_1} C_{n_1,m_1} \int d\vr_1  \int d\vr_3 \tilde f_{n'_1,m'_1}(r_1)Q_{\ell',m'_3}(r_3) \bar V^{sc}_{|m'_3-m_3|}(r_1,r_3)\tilde f_{n_1,m_1}(r_1)Q_{\ell,m_3}(r_3)  \nn \\
&&- \delta_{m'_1,m_1}O^{(m_1)}_{n'_1,n_1} C_{n'_2,m'_2} C_{n_2,m_2}\Big[ \int d\vr_2  \int d\vr_3 \tilde f_{n'_2,m'_2}(r_2)Q_{\ell',m'_3}(r_3) \bar V^{sc}_{|m'_3-m_3|}(r_2,r_3)\tilde f_{n_2,m_2}(r_2)Q_{\ell,m_3}(r_3) \nn \\
&&- \int  d\vr_2\int d\vr_3 \tilde f_{n'_2,m'_2}(r_2)Q_{\ell,m_3}(r_2) \bar V^{sc}_{|m_2+m'_3|}(r_2,r_3) Q_{\ell',m'_3}(r_3)\tilde f_{n_2,m_2}(r_3) \Big] \, ,
\eea
where $m_2=-(m_1+m_3), m'_2=-(m'_1+m'_3)$, and
\be
\lan \ell',m_3|V_{eh}|\ell,m_3 \ran =   \int d{\bf r}  Q_{\ell',m_3}(r) Q_{\ell,m_3}(r) \tilde V^{sc}(r).  \, \label{PE2}
\ee
$\lan \ell',m_3|V_{eh}|\ell,m_3 \ran$ can also be evaluated in momentum space as in Eq.~(\ref{Veh}) but with $\tilde V^{sc}_0(q',q)$ replaced by $\tilde V^{sc}_{m_3}(q',q)$.
\mbox{}

$\bullet$  {\bf Unpolarized Fermi sea}

\mbox{}

For unpolarized FS ($N_+=N_-=N/2$), we must also add  the contribution of the basis states $|n_1,n_2,\ell,m_1,m_3,-1/2\ran$ for all possible single electron-hole pair excitations resulting in a spin-$(-1/2)$ FS hole. Here we have two electrons  of the same spin above the Fermi surface. Thus, it important to keep track on the anticommutation of the two-electron states. It is more convenient to use an orthogonal set of one-particle basis functions for both electrons (e1 and e2) via the Gram-Schmidt process. After the orthonormalization process,the basis states are denoted by $|\tilde n_1,\tilde n_2,\ell,m_1,m_3,-1/2\ran$.
We write the orthonormalized basis functions as
\be \lan \vr |\tilde n, m \ran = \tilde g_{\tilde n,m}(r)e^{im\varphi}. \ee
Here we impose the constraint $\tilde n_1<\tilde n_2$ for $m_1=m_2$ in the basis set due to the Pauli exclusion principle.  Since thee basis states only couple to the exciton states with a frozen FS, the additional matrix elements are
\bea
\lan \Psi^{(e)}_{j'}|H|\tilde n_1,\tilde n_2,\ell,m_1,m_3,-\frac 1 2\ran = \sum_{\tilde n'_1} x^{(j')}_{\tilde n'_1,0}
\Big[\int d\vr_1 \tilde g_{\tilde n'_1,0}(r_1)\tilde g_{\tilde n_1,m_1}(r_1) \int d\vr_2 \tilde g_{\tilde n_2,m_2}(r_2) Q_{\ell,m_3}(r_2) \bar V^{sc}_{|m_1|}(r_1,r_2)  \nn \\
 -\int d\vr_1 \tilde g_{\tilde n'_1,0}(r_1)\tilde g_{\tilde n_2,m_2}(r_1) \int d\vr_2 \tilde g_{\tilde n_1,m_1}(r_2) Q_{\ell,m_3}(r_2)  \bar V^{sc}_{|m_2|}(r_1,r_2)\Big] \nn \\
+\Big[\delta_{m_1,0} x^{(j')}_{\tilde n_1,0}\lan \tilde n_2,m_3|\tilde V_{eh}|\ell,m_3\ran -\delta_{m_2,0}  x^{(j')}_{\tilde n_2,0}\lan \tilde n_1,m_3|\tilde V_{eh}|\ell,m_3\ran \Big]
 \,
\eea
and
\bea  \lan \tilde n'_1,\tilde n'_2,\ell',m'_1,m'_3,-\frac 1 2| H |\tilde n_1,\tilde n_2,\ell,m_1,m_3,-\frac 1 2\ran =\delta_{m'_1,m_1} \delta_{m'_3,m_3} \Big\{ \big[K^{(m_1)}_{\tilde n'_1,\tilde n_1}+\lan \tilde n'_1,m_1|\tilde V_{eh}|\tilde n_1,m_1 \ran \big] \delta_{\tilde n'_2,\tilde n_2}\delta_{\ell',\ell} \nn \\
+ \delta_{\tilde n'_1,\tilde n_1}\big[K^{(m_2)}_{\tilde n'_2,\tilde n_2}+\lan \tilde n'_2,m_2|\tilde V_{eh}|\tilde n_2,m_2 \ran \big]\delta_{\ell',\ell}
+\delta_{\tilde n'_1,\tilde n_1}\delta_{\tilde n'_2,\tilde n_2}\big[- \tilde K_{\ell',\ell}+\lan \ell',m_3|\tilde V_{eh}|\ell,m_3 \ran \big] \Big\}     \nn \\
+  \delta_{m'_3,m_3}\delta_{\ell',\ell} \Big[ \int d\vr_1 \tilde g_{\tilde n'_1,m'_1}(r_1)\tilde g_{\tilde n_1,m_1}(r_1) \int  d\vr_2 \tilde g_{\tilde n'_2,m'_2}(r_2)\tilde g_{\tilde n_2,m_2}(r_2) \bar  V^{sc}_{|m_1-m'_1|}(r_1,r_2)\nn \\
- \int d\vr_1 \tilde g_{\tilde n'_1,m'_1}(r_1)\tilde  f_{\tilde n_2,m_2}(r_1) \int d\vr_2 \tilde g_{\tilde n'_2,m'_2}(r_2)\tilde  f_{\tilde n_1,m_1}(r_2) \bar  V^{sc}_{|m_2-m'_1|}(r_1,r_2) \Big] \nn \\
-  \delta_{m'_2,m_2}\delta_{n'_2,n_2}\Big[ \int d\vr_1  \int d\vr_3 \tilde g_{\tilde n'_1,m'_1}(r_1)Q_{\ell',m'_3}(r_3) \bar V^{sc}_{|m'_3-m_3|}(r_1,r_3)\tilde g_{\tilde n_1,m_1}(r_1)Q_{\ell,m_3}(r_3)  \nn \\
-  \int  d\vr_1 \int d\vr_3 \tilde g_{\tilde n'_1,m'_1}(r_1)Q_{\ell',m'_3}(r_3) \bar V^{sc}_{|m'_3-m_1|}(r_1,r_3)\tilde g_{\tilde n_1,m_1}(r_3)Q_{\ell,m_3}(r_1)\Big] \nn \\
+  \delta_{m'_2,m_1}\delta_{\tilde n'_2,\tilde n_1}\Big[ \int d\vr_1  \int d\vr_3 \tilde g_{\tilde n'_1,m'_1}(r_1)Q_{\ell',m'_3}(r_3) \bar V^{sc}_{|m'_3-m_3|}(r_1,r_3)\tilde g_{\tilde n_2,m_2}(r_1)Q_{\ell,m_3}(r_3)  \nn \\
-  \int  d\vr_1 \int d\vr_3 \tilde g_{\tilde n'_1,m'_1}(r_1)Q_{\ell',m'_3}(r_3) \bar V^{sc}_{|m'_3-m_2|}(r_1,r_3)\tilde g_{\tilde n_2,m_2}(r_3)Q_{\ell,m_3}(r_1)\Big] \nn \\
- \delta_{m'_1,m_1}\delta_{\tilde n'_1,\tilde n_1} \Big[ \int d\vr_2 \int d\vr_3  \tilde g_{\tilde n'_2,m'_2}(r_2)Q_{\ell',m'_3}(r_3) \bar V^{sc}_{|m'_3-m_3|}(r_2,r_3)\tilde g_{\tilde n_2,m_2}(r_2)Q_{\ell,m_3}(r_3) \nn \\
- \int  d\vr_2\int d\vr_3 \tilde g_{\tilde n'_2,m'_2}(r_2)  Q_{\ell',m'_3}(r_3)\bar V^{sc}_{|m'_3-m_2|}(r_2,r_3)Q_{\ell,m_3}(r_2)\tilde g_{\tilde n_2,m_2}(r_3) \Big]  \nn \\
+ \delta_{m'_1,m_2}\delta_{\tilde n'_1,\tilde n_2} \Big[ \int d\vr_2 \int d\vr_3  \tilde g_{\tilde n'_2,m'_2}(r_2)Q_{\ell',m'_3}(r_3) \bar V^{sc}_{|m'_3-m_3|}(r_2,r_3)\tilde g_{\tilde n_1,m_1}(r_2)Q_{\ell,m_3}(r_3) \nn \\
- \int  d\vr_2\int d\vr_3 \tilde g_{\tilde n'_2,m'_2}(r_2)  Q_{\ell',m'_3}(r_3)\bar V^{sc}_{|m'_3-m_1|}(r_2,r_3)Q_{\ell,m_3}(r_2)\tilde g_{\tilde n_1,m_1}(r_3) \Big]
\eea
where $m'_2=-(m'_1+m'_3)$ and $m_2=-(m_1+m_3)$.

\end{widetext}

\mbox{}\\{}*{yiachang@gate.sinica.edu.tw}

\end{document}